\documentclass{gSCS2e}
\usepackage{fontenc}
\usepackage{color}
\usepackage{amsmath}

\usepackage{array}
\usepackage{amsfonts}
\usepackage{amssymb}
\usepackage{euscript}
\usepackage{delarray}
\usepackage{latexsym}
\usepackage{enumerate}
\usepackage{mathtools}
\usepackage{dsfont}
\usepackage{fancybox}

\usepackage{graphicx}
 \usepackage{subfigure}
\usepackage{multirow}
\usepackage{multicol}

\theoremstyle{plain}
\newtheorem{theorem}{Theorem}[section]

\newtheorem{lemma}[theorem]{Lemma}
\newtheorem{proposition}[theorem]{Proposition}

\theoremstyle{remark}
\newtheorem{remark}[theorem]{Remark}
\newtheorem{definition}[theorem]{Definition}

 \newenvironment{proc}[2]{
 
 {} 
 \bigskip 
 \emph{\textbf{ #1 Procedure}}
 \itshape{ #2}
 }

\newcounter{mescond} 
\renewcommand{\themescond}{C.\arabic{mescond}} 

\newenvironment{cond}[1]{ 
\addtocounter{mescond}{0}
\refstepcounter{mescond} 
{}
\vspace{0.5cm}
\begin{center}
 \framebox[1.05\width]{\textit{#1} }\quad (\themescond) 
\end{center}
\vspace{0.5cm}
}

\newlength{\eqboxstorage}
\newcommand{\eqbox}[1]{
  \setlength{\eqboxstorage}{\fboxsep}
  \setlength{\fboxsep}{6pt}
  \boxed{#1}
  \setlength{\fboxsep}{\eqboxstorage}
}

\newcommand\beq{\begin{equation}}
\newcommand\eeq{\end{equation}}

\newcommand\beqe{\begin{equation*}}
\newcommand\eeqe{\end{equation*}}

\newcommand\barr{\begin{array}}
\newcommand\earr{\end{array}}

\newcommand{\id}{
\text{I}
}

 \newcommand{\argmin}{\operatornamewithlimits{Argmin}}

\newcommand{\alsur}[2]{ 
#1\overset{ a.s.}{\rightarrow}#2
}

\newcommand{\normemp}[1]{
\left\| #1 \right\|_{\text{\tiny n}}
}

\newcommand{\psemp}[2]{
\langle #1,#2 \rangle_{\text{\tiny n}}
}

\newcommand{\abs}[1]{
\left\lvert #1 \right\rvert
}

\newcommand{\norm}[1]{
\left\| #1 \right\|
}

\DeclarePairedDelimiter\psth{\langle}{\rangle}

\newcommand{\eucl}[1]{
\left\| #1 \right\|_{\text{\tiny 2}}
}

\newcommand{\disc}[2]{
[#1 : #2]
}

\newcommand{\normmat}[1]{
\left\lvert\!\left\lvert\!\left\lvert #1 \right\rvert\!\right\rvert\!\right\rvert_{\text{\tiny 2}}
}

\newcommand{\normmatf}[1]{
\left\lvert\!\left\lvert\!\left\lvert #1 \right\rvert\!\right\rvert\!\right\rvert_{\text{\tiny F}}
}

\newcommand{\gener}[1]{
\mbox{Span}\left\{ #1 \right\}
}

\newcommand{\bolds}[1]{
\boldsymbol{#1}}

\newcommand{\pxae}{
P_{\mathbf X}-\text{a.e.}}

\newcommand{\cp}[1]{\underset{#1}{\times}}

\newcommand{\h}{H_u^{0,L}}
\newcommand{\hath}{\hat H_u^{0,L}}

\begin{document}


\articletype{Special Issue: SAMO-MASCOT}

\title{{\itshape Journal of Statistical Computation and Simulation} \break Generalized Sobol sensitivity indices for dependent variables: numerical methods}

\author{G. Chastaing$^{a}$ $^{\ast}$, \thanks{$^\ast$ Corresponding author. Email: gal.chastaing@gmail.com
\vspace{6pt}} F. Gamboa$^{b}$  \vspace{6pt} and C. Prieur$^{a}$\\\vspace{6pt}  $^{a}${\em{Universit\'e de Grenoble, LJK/MOISE BP53 38041 Grenoble cedex, France}};\\
$^{b}${\em{Universit\'e Paul Sabatier, IMT-EPS,  118, Route de Narbonne, 31062 Toulouse  Cedex 9, France}}\\\received{v1.0 released October 2013} }

\maketitle

\begin{abstract}
The hierarchically orthogonal functional decomposition of any measurable function $\eta$ of a random vector $\mathbf X=(X_1,\cdots,X_p)$ consists in decomposing $\eta(\mathbf X)$ into a sum of increasing dimension functions depending only on a subvector of $\mathbf X$. Even when $X_1,\cdots, X_p$ are assumed to be dependent, this decomposition is unique if the components are hierarchically orthogonal. That is, two of the components are orthogonal whenever all the variables involved in one of the summands are a subset of the variables involved in the other. Setting $Y=\eta(\mathbf X)$, this decomposition leads to the definition of generalized sensitivity indices able to quantify the uncertainty of $Y$ due to each dependent input in $\mathbf X$~\cite{chastaing}. In this paper, a numerical method is developed to identify the component functions of the decomposition using the hierarchical orthogonality property. Furthermore, the asymptotic properties of the components estimation is studied, as well as the numerical estimation of the generalized sensitivity indices of a toy model. 
Lastly, the method is applied to a model arising from a real-world problem.
\medskip

\medskip

\begin{classcode}62G08, 62H99\end{classcode}\medskip

\end{abstract}


\section{Introduction}\label{sect1}
In a nonlinear regression model, input parameters can be affected by many sources of uncertainty. 
The objective of global sensitivity analysis is to identify and to rank the input variables that drive the uncertainty of the model output. The most popular methods are the variance-based ones~\cite{saltelli}. Among them, the Sobol indices are widely used~\cite{sobol}. This last method relies on the assumption that the inputs are independent. Under this assumption, Hoeffding~\cite{hoef} shows that the model output can be uniquely cast as a sum of increasing dimension functions, where the integrals of every summand with respect to any of its own variables must be zero. A consequence of these conditions is that all summands of the decomposition are mutually orthogonal.
Later, Sobol applies the latter decomposition to sensitivity analysis~\cite{sobol}. It results that the global variance can be decomposed as a sum of partial variances that quantify the sensitivity of a set of inputs on the model response. 

However, for models featuring dependent inputs, the use of Sobol indices may lead to a wrong interpretation because the sensitivity induced by the dependence between two factors is implicitly included in their Sobol indices. To handle this problem, a straightforward solution consists in computing Sobol sensitivity indices for independent groups of dependent variables. First introduced by Sobol~\cite{sobol}, this idea is exploited in practice by Jacques \textit{et al.}~\cite{jacques}. Nevertheless, this technique implies to work with models having several independent groups of inputs. Furthermore, it does not allow to quantify the individual contribution of each input. A different way to deal with this issue has been initiated by Borgonovo \textit{et al.}~\cite{borgonovo,borgonovo2}. These authors define a new measure based on the joint distribution of $(Y, \mathbf X)$. The new sensitivity indicator of an input $X_i$ measures the shift between the output distribution and the same distribution conditionally to $X_i$. This moment free index has many properties and has been applied to some real applications~\cite{borgonovo3,borgonovo4}. However, the dependence issue remains unsolved as we do not know how the conditional distribution is distorted by the dependence, and how it impacts the sensitivity index. Another idea is to use the Gram-Schmidt orthogonalization procedure. In an early work, Bedford~\cite{bedford} suggests to orthogonalize the conditional expectations and then to use the usual variance decomposition on this new orthogonal collection. Further, the Monte Carlo simulation is used to compute the indices. 
Following this approach,the Gram-Schmidt process is exploited by Mara \textit{et al.}~\cite{mara}, before performing a polynomial regression to approximate the model. In both papers, the decorrelation method depends on the ordering of the variables, making the procedure computationally expensive and difficult to interpret. \\
Following the construction of Sobol indices previously exposed, Xu \textit{et al.}~\cite{xu} propose to decompose the partial variance of an input into a correlated and an uncorrelated contribution in the context of linear models. This last work has been later extended by Li \textit{et al.} with the concept of HDMR~\cite{li,hdmr}. In~\cite{li}, the authors suggest to reconstruct the model function using classical bases (polynomials, splines,...), then to deduce the decomposition of the response variance as a sum of partial variances and covariances. Instead of a classical basis, Caniou \textit{et al.}~\cite{caniou} use a polynomial chaos expansion to approximate the initial model, and the copula theory to model the dependence structure~\cite{nelsen}. Thus, in all these papers, the authors choose a type of model reconstruction before proceeding to the splitting of the response variance.\\ 

In a previous paper~\cite{chastaing}, we revisited the Hoeffding decomposition in a different way, leading to a new definition of the functional decomposition in the case of dependent inputs. Inspired by the pioneering work of Stone~\cite{stone} and Hooker~\cite{hooker}, we showed, under a weak assumption on the inputs distribution, that any model function can be decomposed into a sum of hierarchically orthogonal component functions. Hierarchical orthogonality means that two of these summands are orthogonal whenever all variables included in one of the components are also involved in the other. The decomposition leads to generalized Sobol sensitivity indices able to quantify the uncertainty induced by the dependent model inputs.

The goal of this paper is to complement the work done in~\cite{chastaing} by providing an efficient numerical method for the estimation of the generalized Sobol sensitivity indices. In our previous paper~\cite{chastaing}, we have proposed a statistical procedure based on projection operators to identify the components of the hierarchically orthogonal functional decomposition (HOFD).
The method consists in projecting the model output onto constrained spaces to obtain a functional linear system. 
The numerical resolution of these systems relies on an iterative scheme that requires to estimate conditional expectations at each step. 
This method is well tailored for independent pairs of dependent variables models. However, it is difficult to apply to more general models because of its computational cost. 
Hooker~\cite{hooker} has also worked on the estimation of the HOFD components. This author studies the component estimation via a minimization problem under constraints using a sample grid. 
In general, this procedure is also quite computationally demanding. Moreover, it requires to get a prior on the inputs distribution at each evaluation point, or, at least, to be able to estimate them properly. 
In a recent article, Li {\it{et al.}}~\cite{li2} reconsider Hooker's work and also identify the HOFD components by a least-squares method. These last authors propose to approximate these components expanded on a suitable basis. 
They bypass some technical problem of degenerate design matrix by using a continuous descent technique~\cite{li2010}.

In this paper, we propose an alternative to directly construct a hierarchical orthogonal basis. Inspired by the usual Gram-Schmidt algorithm, the procedure consists in recursively constructing for each component a multidimensional basis that satisfies the hierarchical orthogonal conditions. This procedure will be referred as the Hierarchically Orthogonal Gram-Schmidt (HOGS) procedure. Then, each component of the decomposition can be properly estimated by a linear combination of this basis. The coefficients are then estimated by the usual least-squares method. Thanks to the HOGS Procedure, we show that the design matrix has full rank, so the minimization problem admits a unique and explicit solution. Furthermore, we study the asymptotic properties of the estimated components. 
Nevertheless, the practical estimation of the one-by-one component suffers from the curse of dimensionality when using the ordinary least-squares estimation. To handle this problem, we propose to estimate parameters of the model using variable selection methods. 
Two usual algorithms are briefly presented, and are adapted to our method. Moreover, the HOGS Procedure coupled with these algorithms is experimented on numerical examples. 

The paper is organized as follows.
In Section \ref{sect2}, we give and discuss the general results related to the HOFD. We remind Conditions (\ref{c1}) and (\ref{c2}) under which the HOFD is available. Further, a definition of generalized Sobol sensitivity indices is derived and discussed. Section \ref{sect3} is devoted to the HOGS Procedure. We introduce the appropriate notation, and outline the procedure in detail. In Section \ref{sect4}, we adapt the least-squares estimation to our problem, and we point out the curse of dimensionality. In addition, the use of a penalized minimization scheme is considered in order to perform model selection.
The last part of Section \ref{sect4} is devoted to the consistency of the HOGS Procedure.
In Section \ref{sectappli}, we present numerical applications. The first two examples are toy function. The objective here is to show the efficiency of the HOGS Procedure coupled with variable selection methods to estimate the sensitivity indices. The last example is an industrial case study. The objective is to detect the inputs that have the strongest impact on a tank distortion.   


\section{Generalized Sobol sensitivity indices}\label{sect2}
Functional ANOVA models are specified by a sum of functions depending on an increasing number of variables.
A functional ANOVA model is said to be additive if only main effects are included in the model. It is said to be saturated if all interaction terms are included in the model. The existence and the uniqueness of such a decomposition is ensured by some identifiability constraints. 
When the inputs are independent, any squared-integrable model function can be exactly represented by a saturated ANOVA model with pairwise orthogonal components. As a result, the contribution of any group of variables to the model response is measured by the Sobol index, bounded between $0$ and $1$. Moreover, the sum of all the Sobol indices is equal to $1$~\cite{sobol}. The use of such an index is not excluded in the context of dependent inputs, but the information conveyed by the Sobol indices is redundant, and may lead to a wrong interpretation of the sensitivity in the model. In this section, we remind the main results established in Chastaing \textit{et al.}~\cite{chastaing} for possibly non-independent inputs. In this case, the saturated ANOVA decomposition holds with weaker identifiability constraints than for the independent case. This leads to a generalization of the Sobol indices that is well suited to perform global sensitivity analysis when the inputs are not necessarily independent.\\
First, we remind the general context and notation. 
The last part is dedicated to the generalization of the Hoeffding-Sobol decomposition when inputs are potentially dependent. The definition of the generalized sensitivity indices is introduced in the following.

\subsection{First settings}\label{sect21}

\noindent Consider a measurable function $\eta$ of a random vector $\mathbf X=(X_1,\cdots,X_p) \in \mathbb R^p$, $p\geq 1$, and let $Y$ be the real-valued response variable defined as
 $$
 \begin{array}{cc}
 Y :&
\begin{array}{ccc}
 (\mathbb{R}^p,\mathcal{B}(\mathbb{R}^p),P_{\mathbf{X}}) &\rightarrow & (\mathbb{R},\mathcal{B}(\mathbb{R})) \\
 \mathbf X & \mapsto & \eta(\mathbf X)
\end{array}
\end{array}
$$

 where the joint distribution of $\mathbf X$ is denoted by $P_{\mathbf{X}}$. For a $\sigma$\textendash finite measure $\nu$ on $(\mathbb{R}^p,\mathcal{B}(\mathbb{R}^p))$, we assume that $P_{\mathbf X} << \nu$ and that $\mathbf X$ admits a density $p_{\mathbf{X}}$ with respect to $\nu$, that is $p_{\mathbf{X}}=\dfrac{dP_{\mathbf{X}}}{d\nu}$.\\ 
 Also, we assume that $\eta \in L^2_{\mathbb R}(\mathbb R^p,\mathcal{B}(\mathbb{R}^p),P_{\mathbf{X}})$. As usual, we define the inner product $\langle \cdot,\cdot \rangle$ and the norm $\|\cdot \|$ of the Hilbert space $L^2_{\mathbb R}(\mathbb R^p,\mathcal{B}(\mathbb{R}^p),P_{\mathbf{X}})$ as
\begin{equation*}
\begin{array}{l}
 \langle h_1,h_2\rangle=\int h_1(\mathbf x)h_2(\mathbf x) p_{\mathbf{X}} d\nu(\mathbf x) = \mathbb E(h_1(\mathbf X)h_2(\mathbf X)), ~~h_1,h_2 \in  L^2_{\mathbb R}(\mathbb R^p,\mathcal{B}(\mathbb{R}^p),P_{\mathbf{X}})\\
 \\
 \|h\|^2=\langle h,h\rangle=\mathbb E(h(\mathbf X)^2), \quad h\in L^2_{\mathbb R}(\mathbb R^p,\mathcal{B}(\mathbb{R}^p),P_{\mathbf{X}})
\end{array}
\end{equation*}
Here $\mathbb E(\cdot)$  denotes the expectation.
Further, $V(\cdot)=\mathbb E[(\cdot-\mathbb E(\cdot))^2]$ denotes the variance, 
and $\mbox{Cov}(\cdot,\ast)=\mathbb E[(\cdot-\mathbb E(\cdot))(\ast-\mathbb E(\ast))]$ the covariance.\\
Let us denote $\disc{1}{k}:=\{1,2,\cdots,k\}$, $\forall~k \in \mathbb{N}^*$, and let $S$ be the collection of all subsets of $\disc{1}{p}$. As misuse of notation, we will denote the sets $\{i\}$ by $i$, and $\{ij\}$ by $ij$.
For $u \in S$ with $u=\{u_1,\cdots,u_k\}$, we denote the cardinality of $u$ by $|u|=k$ and the corresponding random subvector by $\mathbf {X_u}:=(X_{u_1},\cdots,X_{u_k})$. Conventionally, if $u=\emptyset$, $|u|=0$, and $X_{\emptyset}=1$. Also, we denote by $\mathbf{X_{-u}}$ the complementary vector of $\mathbf{X_u}$ (that is, $-u$ is the complementary set of $u$). The marginal density of $\mathbf{X_u}$ (respectively $\mathbf{X_{-u}}$) is denoted by $p_{\mathbf{X_u}}$ (resp. $p_{\mathbf{X_{-u}}}$).\\

 Further, the mathematical structure of the functional ANOVA models is defined through subspaces $(H_u)_{u\in S}$ and $(H_u^0)_{u \in S}$ of $L^2_{\mathbb R}(\mathbb R^p,\mathcal{B}(\mathbb{R}^p),P_{\mathbf{X}})$. $H_{\emptyset}\equiv H_{\emptyset}^0$ denotes the space of constant functions. For $u\in S\setminus \{\emptyset\}$, $H_u$ is the space of square-integrable functions that depend only on $\mathbf{X_u}$. The space $H_u^0$ is defined as:
\[
 H_u^0=\left\{ h_u\in H_u, ~\langle h_u,h_v\rangle=0, \forall~v \subset u, \forall~h_v \in H_v^0 \right\}=H_u \cap \left(\sum_{v \subset u} H_v^0\right )^{\perp},
\]
where $\subset$ denotes the strict inclusion, that is $A \subset B \Rightarrow A \cap B \neq B$. Further, $\subseteq$ will denote the inclusion when equality is possible.\\



\subsection{Generalized Sobol sensitivity indices}

Let us suppose that
\cond{
$
 \begin{array}{lc}
&P_{\mathbf{X}} << \nu \\
\textrm{where}&\\
&\nu(dx)=\nu_1(dx_1) \otimes \cdots \otimes \nu_p(dx_p)
\end{array}
$
}\label{c1}

Our main assumption is :
\cond{
$
\begin{array}{lll}
\exists~ 0<M\leq 1 , ~\forall~u \in S\setminus \{\emptyset\},&  p_{\mathbf{X}}\geq M \cdot p_{\mathbf{X_u}}p_{\mathbf{X_{-u}}} \quad \nu\textrm{-a.e.} 
\end{array}
$
 }\label{c2} 

Under these conditions, the following result states a general decomposition of $\eta$ as a saturated functional ANOVA model, under the specific conditions of the spaces $H_u^0$ (defined in Section \ref{sect21}),

\begin{theorem}\label{theor1}
Let $\eta$ be any function in $L^2_{\mathbb R}(\mathbb R^p,\mathcal{B}(\mathbb{R}^p),P_{\mathbf{X}})$. Then, under (\ref{c1}) and (\ref{c2}), there exist unique functions $(\eta_0,\eta_1,\cdots,\eta_{\{1,\cdots,p\}}) \in H_{\emptyset}\times H_1^0 \times \cdots H_{\{1,\cdots,p\}}^0$ such that the following equality holds :
\begin{equation}\label{equ2}
\begin{array}{lll}
  \eta(\mathbf X)&=&\displaystyle{\eta_0+\sum_{i=1}^p \eta_i(X_i)+\sum_{1\leq i<j\leq p}\eta_{ij}(X_i,X_j)+\cdots+\eta_{\{1,\cdots,p\}}(\mathbf X)}\\
&=&\displaystyle{\sum_{u \in S} \eta_u(\mathbf{X_u})} \quad\textrm{a.e.}
\end{array}
\end{equation}
\end{theorem}

To get a better understanding of Theorem \ref{theor1}, the reader could refer to its proof and further explanations in \cite{chastaing}. Notice that, unlike the Sobol decomposition with independent inputs, the component functions of (\ref{equ2}) are hierarchically orthogonal, and no more mutually orthogonal. Thus, from now on, the obtained decomposition (\ref{equ2}) will be abbreviated HOFD (for Hierarchically Orthogonal Functional Decomposition). Also, as mentioned in \cite{chastaing}, the HOFD is said to be a generalized decomposition because it turns out to be the usual functional ANOVA decomposition when inputs are independent. \\
The general decomposition of the output $Y=\eta(\mathbf X)$ given in Theorem \ref{theor1} allows one to decompose the global variance as a simplified sum of covariance terms. Further below, we define the generalized sensitivity indices able to measure the contribution of any group of inputs in the model when inputs can be dependent :

\begin{definition}\label{def1}
 The sensitivity index $S_u$ of order $|u|$ measuring the contribution of $\mathbf{X_u}$ to the model response is given by :
 \begin{equation}\label{eq10}
\eqbox{
S_u=\dfrac{V(\eta_u(\mathbf{X_u}))+ \sum_{\substack{v\in S\setminus \{\emptyset\}\\ u\cap v\neq \{u,v\}}} \mbox{Cov} (\eta_u(\mathbf{X_u}),\eta_v(\mathbf{X_v}))}{V(Y)}
} 
\end{equation}
More specifically, the first order sensitivity index $S_i$ is given by :
\begin{equation}\label{eq11}
 \eqbox{
S_i=\dfrac{V(\eta_i(X_i))+ \sum_{\substack{v \neq \emptyset \\ i \not\in v }} \text{Cov} (\eta_i(X_i),\eta_v(\mathbf{X_v}))}{V(Y)}
} 
\end{equation}
\end{definition}
These indices are called generalized Sobol sensitivity indices because if all inputs are independent,  it can be shown that $\mbox{Cov}(\eta_u,\eta_v)=0$, $\forall~u \neq v$~\cite{chastaing}. Therefore, in the independent case, the index given by (\ref{eq10}) or (\ref{eq11}) coincide with the usual Sobol index.

\begin{proposition}\label{pro2}
Under (\ref{c1}) and (\ref{c2}), the sensitivity 
indices $S_u$ previously defined sum to $1$, i.e. $ \sum_{ u \in S\setminus\{\emptyset\}} S_u=1$.
%
\end{proposition}

\paragraph*{\textbf{Interpretation of the sensitivity indices}}
The indices' interpretation is not obvious, as they are no more bounded and they can even be negative. We provide here an interpretation of the first order sensitivity index $S_i$, split into two parts:
\begin{equation*}
S_i=\underbrace{\dfrac{V(\eta_i(X_i))}{V(Y)}}_{\mathrm{VS}_i}+ 
\underbrace{\dfrac{
\sum_{\substack{v \neq \emptyset \\ i \not\in v }} \text{Cov} (\eta_i(X_i),\eta_v(\mathbf{X_v}))}{V(Y)}}_{\mathrm{CoVS}_i}.
\end{equation*}
The first part, $\mathrm{VS}_i$, could be identified as the full contribution of $X_i$, whereas the second part, $\mathrm{CoVS}_i$, could be interpreted as the contribution induced by the dependence with the other terms of the decomposition. Thus,  $\mathrm{CoVS}_i$ would play the role of compensation. We detail here this interpretation, and we distinguish five cases, represented and explained further below.

\begin{minipage}[c]{0.5\textwidth}
\centering
 \includegraphics[width=0.9\textwidth]{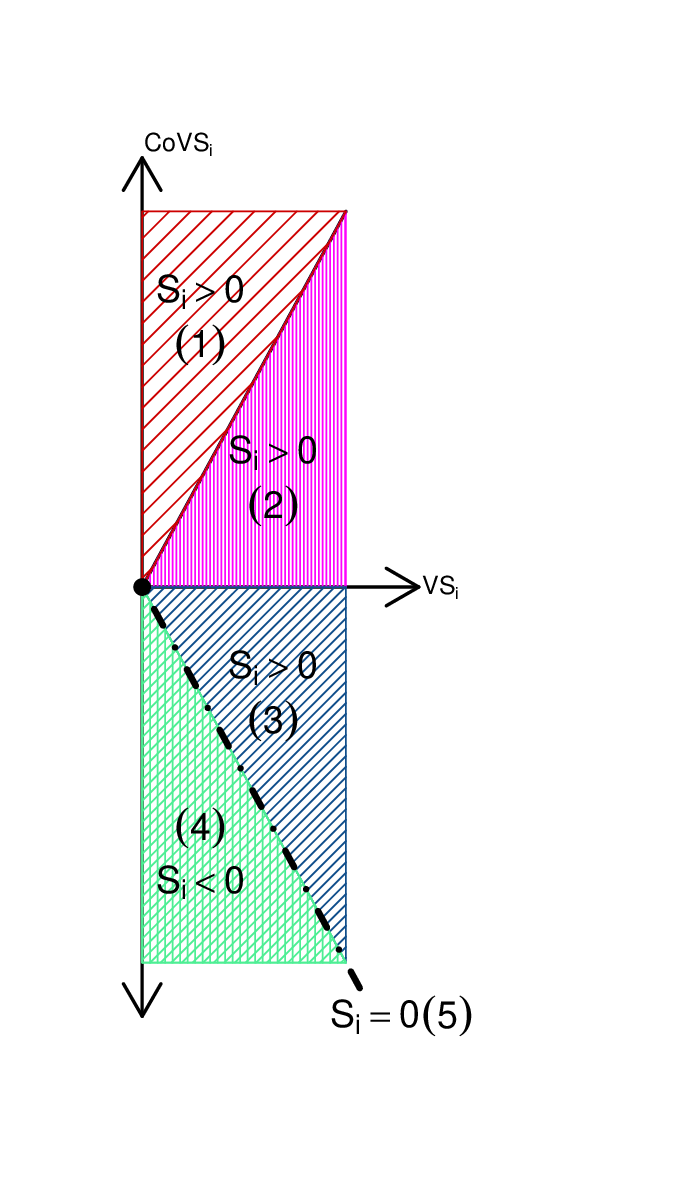}
\end{minipage}
\begin{minipage}[c]{0.5\textwidth}
\textbf{Case (1).} The full contribution of $X_i$ is not important, but the uncertainty is induced by the dependence. Thus, $X_i$ has an influence through its dependence with the other variables.\\
\textbf{Case (2).} The full contribution of $X_i$ is important, and the induced contribution is lower. In this case, $X_i$ has a strong influence.\\
\textbf{Case (3).} The uncertainty of the sole $X_i$ is important, but weakened by the contribution induced by the dependence.\\
\textbf{Case (4).} The influence of $X_i$ is not direct, as it comes from its dependencies with other factors. This influence, obtained by negative covariances, is more important than the full contribution, so $X_i$ may not be so significant.\\
\textbf{Case (5).} The case $\mathrm{VS}_i=\mathrm{CoVS}_i=0$ means that $X_i$ is not contributive to the global variability. However, if $\mathrm{VS}_i=-\mathrm{CoVS}_i\neq 0$, the dependence makes a perfect compensation. 
\end{minipage}
The interpretation is subject to the index splitting, so it may have an impact on the conclusion drawn in sensitivity analysis. This impact has to be carefully considered and its study remains an open problem.

From now on, the next part is dedicated to the practical use of these indices. The analytical formulation of the HOFD components is rarely available in realistic applications. Indeed, their computation requires to know the mathematical form of $\eta$ and the distribution of the input parameters. It also implies to search for components in a space constrained by very specific orthogonality conditions. Efficient numerical methods has then to be developed to estimate the generalized sensitivity indices.
The following section is devoted to an original estimation scheme of the HOFD components based on two tools: extended bases~\cite{li2} and penalized regression~\cite{hastie}. 


\section{The hierarchically orthogonal Gram-Schmidt procedure }\label{sect3}

In Section \ref{sect2}, we show that each component $\eta_u$ belongs to a subspace $H_u^0$ of $L^2(\mathbb R^p, \mathcal B(\mathbb R^p), P_{\mathbf X})$. Thus, to estimate $\eta_u$, the most natural approach is to construct a good approximation space of $H_u^0$. In addition,
 we have seen that the generalized sensitivity indices are defined for any type of reference measures $(\nu_i)_{i \in \disc{1}{p}}$. From now and until the end, we will assume that $\nu_i$, $\forall~i\in \disc{1}{p}$, are diffuse measures. Indeed, the non diffuse measures raise additional issues in the results developed further that we will not address in this paper.\\

In a Hilbert space, it is usual to call in an orthonormal basis to express any of the space element as a linear combination of these basis. Further below, we will define the finite-dimensional spaces $H_u^L \subset H_u$ and $H_u^{0,L} \subset H_u^0$, $\forall~u \in S$, as linear spans of some orthonormal systems that will be settled later. We denote by $\gener{B}$ the set of all finite linear combination of elements of $B$, also called the linear span of $B$.

Consider, for any $i\in \disc{1}{p}$, a truncated orthonormal system $(\psi_{l_i}^i)_{l_i=0}^{L_i}$ of $L^2(\mathbb R,\mathcal B(\mathbb R),P_{X_i})$, with $L_i\geq 1$. Without loss of generality, we simplify the notation, and we assume that $L_i=L\geq 1$, for all $i \in \disc 1 p$. Also, when there is no confusion, $\psi_{l_i}^i$ is written $\psi_{l_i}$. Moreover, we set $\psi_0=1$.
 For any $u=\{u_1,\cdots,u_k\}\in S$, $\bolds{l_u}=(l_{u_1},\cdots, l_{u_k})\in \disc 1 L ^{|u|}$ is the multi-index  associated with the tensor-product $(\otimes_{i =1}^k \psi_{l_{u_i}})$. To define properly the truncated spaces $H_u^L \subset H_u$, we further assume that

\cond{
$\forall u=\{u_i\}_{i=1}^k\in S, ~\forall~ \bolds{l_u}\in \disc 1 L ^{|u|},~ \int [\prod_{i=1}^k\psi_{l_{u_i}}(x_{u_i})]^2 p_{\mathbf X}d\nu(\mathbf{x}) < +\infty$
}\label{condprod}


\begin{remark} 
A sufficient condition for (\ref{c2}) is to have $0<M_1\leq p_{\mathbf X}\leq M_2$ (see Section 3 of~\cite{chastaing}). In this particular case, it is sufficient to assume that, $\forall i\in \disc 1 p$, $\forall~l_i\in \disc 1 L$, $\int (\prod_{i \in \disc{1}{p}}\psi_{l_i}(x_i))^2 d\nu(\mathbf{x})<+\infty$ to ensure (\ref{condprod}).
\end{remark}

Under (\ref{condprod}), we define, $H_{\emptyset}^L=\gener{1}$. Also, we set, $\forall~i\neq j\in \disc{1}{p}$,
\[
H_i^L= \gener{1,\psi_{l_i}, l_i \in \disc 1 L}.
\]
Further, we write the multivariate spaces $H_u^L$, for $u\in S$, as
\[
H_u^L=\otimes_{i\in u} H_i^L.
\]
%
Then, the dimension of $H_u^L$  is $\mbox{dim}(H_u^L)=(L+1)^{|u|}$. \\

Now, we focus on the construction of the \textit{theoretical} finite-dimensional spaces $(H_u^{0,L})_{u\in S}$, that corresponds to the constrained subspaces of $(H_u^L)_{u\in S}$. Thus, for all $u\in S$, $H_u^{0,L}$ is defined as

%
%

\[
H_u^{0,L}=\left\{ h_u \in H_u^{L}, ~\langle h_u,h_v\rangle=0, ~\forall ~v \subset u, ~\forall ~h_v \in H_v^{0,L}\right\}
\]

Hence, $\mbox{dim}(\h)=\mbox{dim}(H_u^L)-[\sum_{\substack{v \subset u\\ v \neq  \emptyset}}  L^{|v|}+1]=L^{|u|}$.\\

Given an independent and identically distributed $n$-sample $(y^s,\mathbf x^s)_{s=1,\cdots,n}$ from the distribution of $(Y,\mathbf X)$, the \textit{empirical} version $\hath$ of $\h$ is defined as $ \hat H_{\emptyset}^{0,L}=H_{\emptyset}^{L}$, and


\[
 \hat H_{u}^{0,L}=\left\{ g_u \in H_u^L, ~\langle g_u,g_v\rangle_{\text{\tiny n}}=0, \forall~v \subset u, \forall~g_v \in \hat H_{v}^{0,L} \right\},
\]

where $\langle \cdot,\cdot \rangle_{\text{\tiny n}}$ defines the empirical inner product associated with the $n$-sample. The space $\hath$  varies with sample size $n$, but for notational convenience, we suppress the dependence on $n$.\\

The next procedure is an iterative scheme to construct $(\h)_{u\in S}$ and $(\hath)_{u\in S}$ by taking into account their specific properties of orthogonality. This numerical method is referred as the Hierarchically Orthogonal Gram-Schmidt (HOGS) procedure.


\begin{proc}{Hierarchically Orthogonal Gram-Schmidt}{}
\begin{enumerate}\label{proc1}
\item \label{step0} \textbf{Initialization.} For any $i \in \disc{1}{p}$, we use the truncated orthonormal system $(\psi_{l_i})_{l_i=0}^{L}$. Set $\phi_{l_i}=\psi_{l_i}$, $\forall~ l_i \geq 1$ and $$H_i^{0,L}=\gener{\phi_{l_i}, l_i\in \disc 1 L}.$$
 As $(\phi_{l_i})_{l_i=1}^{L}$ is an orthonormal system, any function $h_i\in H_i^{0,L}$ satisfies $\mathbb E(h_i(X_i))=0$.
\item \label{step1} \textbf{Second order interactions.} Let $u=\{i,j\}$ with $i\neq j\in \disc 1 p$. The space $H_u^{0,L}$ is constructed in a recursive way. By Step \ref{step0}, we have $H_i^{0,L}=\gener{\phi_{l_i}, l_i\in \disc 1 L}$ and $H_j^{0,L}=\gener{\phi_{l_j}, l_j\in \disc 1 L}$. For all $\bolds{l_{ij}}=(l_i,l_j)\in \disc 1 L^2$,
\begin{enumerate}
\item set
\begin{equation*}
 \phi_{\bolds{l_{ij}}}(X_i, X_j)=\phi_{l_i}(X_i)\times \phi_{l_j}(X_j)+\sum_{k=1}^{L}\lambda_{k}^i\phi_{k}^i(X_i)+\sum_{k=1}^{L}\lambda_{k}^j\phi_{k}^j(X_j)+C_{\bolds{l_{ij}}}
\end{equation*}
\item The constants $(C_{\bolds{l_{ij}}},(\lambda_{k}^i)_{k=1}^L,(\lambda_{k}^i)_{k=1}^L)$ are determined by resolving the hierarchical orthogonal constraints,
\begin{equation*}
\left\{
 \begin{array}{ll}
  \langle  \phi_{l_{ij}},\phi_{k}^i \rangle=0, & \forall~k\in \disc{1}{L}\\
 \langle  \phi_{l_{ij}},\phi_{k}^j \rangle=0, & \forall~k\in \disc{1}{L}\\
\langle \phi_{l_{ij}},1\rangle =0
 \end{array}
\right.
\end{equation*}
Finally, $H_{ij}^{0,L}=\gener{\phi_{\bolds{l_{ij}}},~\bolds{l_{ij}}\in \disc 1 L^2}$. Each function $h_{ij}\in H_{ij}^{0,L}$ satisfies the constraints imposed to $H_{ij}^0$.
\end{enumerate}
\item \label{step2} \textbf{Higher interactions.} To build a basis $(\phi_{\bolds{l_u}})_{\bolds{l_u}\in \disc{1}{L}^{|u|}}$ of $H_u^{0,L}$, with $u=(u_1,\cdots,u_k)$, we proceed recursively on $k$. By the previous steps of the Procedure, we have at hand, for any $v\in S$ such that $1\leq |v| \leq k-1$, 
\[
H_v^{0,L}=\gener{\phi_{\bolds{l_v}}, \bolds{l_v}\in \disc 1{L}^{|v|}},\quad \text{dim}(H_v^{0,L})=L^{|v|}.
\]
Now, to obtain $\phi_{\bolds{l_u}}$, for all $\bolds{l_u}=(l_{u_1},\cdots,l_{u_k}) \in\disc{1}{L}^{|u|}$, we proceed as follows,
\begin{enumerate}
\item set
\begin{equation}\label{sett}
\phi_{\bolds{l_u}}(\mathbf{X_u})=\prod_{i =1}^k\phi_{l_{u_i}}(X_{u_i})
+\sum_{\substack{v \subset u\\ v \neq \emptyset}}\sum_{\bolds{l_v}\in \disc 1{L}^{|v|}} \lambda_{\bolds{l_v},\bolds{l_u}} \phi_{\bolds{l_v}}(\mathbf {X_v})+C_{\bolds{l_u}}
\end{equation}
\item \label{step222} compute the $(1+\sum_{\substack{v\subset u\\ v\neq \emptyset }} L^{|v|})$ coefficients $(C_{\bolds{l_u}},(\lambda_{\bolds{l_v},\bolds{l_u}})_{\bolds{l_v}\in \disc 1{L}^{|v|},v \subset u})$ by solving
\begin{equation}\label{hortho}
\left\{
\begin{array}{ll}
 \langle \phi_{\bolds{l_u}},\phi_{\bolds{l_v}}\rangle=0,& \forall v\subset u, ~\forall~\bolds{l_v} \in \disc{1}{L}^{|v|}\\
  \langle \phi_{\bolds{l_u}},1\rangle=0.
\end{array}
\right.
\end{equation}
The linear system (\ref{hortho}), with (\ref{sett}), is equivalent to a sparse matrix system of the form $A_{\phi}^u \Lambda^u=D^{\bolds{l_u}}$, when $C_{\bolds{l_u}}$ has been removed. The matrix $A_{\phi}^u$ is a Gramian matrix involving terms $\mathbb E(\Phi_{v_1}(\mathbf{X_ {v_1}}) {}^t\Phi_{v_2}(\mathbf{X_ {v_2}}))_{v_1,v_2\subset u}$, with $(\Phi_{v_i}(\mathbf{X_{v_i}}))_{\bolds{l_{v_i}}}=\phi_{\bolds{l_{v_i}}}(\mathbf{X_{v_i}}), \bolds{l_{v_i}}\in \disc{1}{L}^{|v_i|}$, $i=1,2$. $\Lambda^u$ involves the coefficients $(\lambda_{\bolds{l_v},\bolds{l_u}})_{\bolds{l_v}\in \disc{1}{L}^{|v|},v \subset u}$, and $D^{\bolds{l_u}}$ involves $-\mathbb{E}(\prod_{u_i\in u} \phi_{l_{u_i}}^\cdot \Phi_{v_i})_{v_i\subset u}$. \\
In Lemma \ref{noexist}, we show that $A_{\phi}^u$ is a definite positive matrix, so the system (\ref{hortho}) admits a unique solution.

\end{enumerate}

 Finally, set $H_u^{0,L}=\gener{ \phi_{\bolds{l_u}},~ \bolds{l_u} \in \disc{1}{L}^{|u|}}$.

\end{enumerate}
\end{proc}
 
The construction of $(\hat H_{u}^{0,L})_{u\in S}$ is very similar to the $(H_u^{0,L})_{u\in S}$ one. However, as the spaces $(\hat H_{u}^{0,L})_{u\in S}$ depend on the observed $n$-sample, their construction requires to assume that the sample size $n$ is larger than the size $L$.\\
To build $\hat H_{i}^{0,L}$, $\forall~i\in \disc 1p$, we use the usual Gram-Schmidt procedure on $(\phi_{l_i})_{l_i=1}^{L}$ to get an orthonormal system $(\hat \phi_{l_i})_{l_i=1}^{L}$ with respect to the empirical inner product $\psemp{\cdot}{\cdot}$. To build $(\hat H_{u}^{0,L})_{u\in S, |u|\geq 2}$, we can simply use the HOGS procedure while replacing  $\psth{\cdot,\cdot}$ with $\psemp{\cdot}{\cdot}$.
Finally, we denote 
$$
\hat H_{u}^{0,L}=\gener{\hat \phi_{\bolds{l_u}}, ~\bolds{l_u}\in \disc{1}{L}^{|u|}}\quad \forall~u \in S\setminus \{\emptyset\}.$$ 

%
%

In practice, polynomials or splines basis functions~\cite{droesbeke} will be considered. Also, as we may be faced to expensive models with a fixed budget, the number of observations $n$ may be limited. In this case, and in view of the HOGS procedure, $L$ should be chosen accordingly.\\ 
In the next section, we discuss the practical estimation of the generalized Sobol sensitivity indices using least-squares minimization, and its consistency. We also discuss the curse of dimensionality, and propose some variable selection methods to circumvent it.


\section{Estimation of the generalized sensitivity indices}\label{sect4}

\subsection{Least-Squares estimation}\label{sectlse}

The effects $(\eta_u)_{u\in S}$ in the HOFD (\ref{equ2}) satisfy
\begin {equation}\label{min1}
   (\eta_u)_{u\in S}=\argmin_{\substack{(h_u)_{u\in S}\\ h_u \in H_u^0}} \mathbb E[(Y-\sum_{u \in S} h_u(\mathbf{X_u}))^2]
\end {equation}

Notice that $\eta_0$, the expected value of $Y$, is not involved in the sensitivity indices estimation. Thus, $Y$ is replaced with $\tilde Y:=Y-\mathbb E(Y)$ in (\ref{min1}). 
Also, the residual term $\eta_{\{1,\cdots,p\}}$ is removed from (\ref{min1}) and is estimated afterwards. 
In Section \ref{sect3}, we defined the approximating spaces $\hat H_{u}^{0,L}$ of $H_u^0$, for $u\in S\setminus\{ \emptyset \}$. Thus, the minimization problem (\ref{min1}) may be replaced with its empirical version,
\begin{equation}\label{min2}
\min_{(\beta_{\bolds{l_u}})_{\bolds{l_u},u}} \frac{1}{n}\sum_{s=1}^{n}\left[\tilde y^s-\sum_{\substack{u \subset \disc{1}{p}\\ u\neq \emptyset}}\sum_ {\bolds{l_u}\in \disc{1}{L}^{|u|}}\beta_{\bolds{l_u}}^u\hat \phi_{\bolds{l_u}}(\mathbf{x_u}^{s})\right]^2
\end{equation}
 where $\tilde y^s:=y^s-\bar y$, $\bar y:=\frac{1}{n}\sum_{s=1}^n y^s$, and where every subspace $\hat H_{u}^{0,L}$ is spanned by the basis functions $(\hat \phi_{\bolds{l_u}})_{\bolds{l_u}\in \disc{1}{L}^{|u|}}$ constructed according to the HOGS Procedure of Section \ref{sect3}, where $\psth{\cdot,\cdot}$ is replaced by $\psemp \cdot \cdot$.
The equivalent matrix form of (\ref{min2}) is
\begin{equation}\label{minmatrix}
   \min_{\boldsymbol\beta} \| \mathbb Y-\mathbb X_{\hat \phi} \boldsymbol\beta\|_{\text{\tiny n}}^2
\end{equation}
where $\mathbb Y_s=y^s-\bar y$, $\mathbb X_{\hat \phi}=\begin{pmatrix} \bolds{\hat \phi}_1 & \cdots & \bolds{\hat \phi}_u & \cdots\end{pmatrix} \in \cp{u\in S}\mathcal M_{n,L^{|u|}}(\mathbb R)$, where $\cp{u\in S}\mathcal M_{n,L^{|u|}}(\mathbb R)$ denotes the cartesian product of real entries matrices with $n$ rows and $L^{|u|}$ columns.

For $u\in S$, $(\bolds{\hat \phi}_u)_{s,\bolds{l_u}}=\hat \phi_{\bolds{l_u}}(\mathbf {x_u}^s)$, $(\bolds\beta)_{\bolds{l_u},u}=\beta_{\bolds{l_u}}^u$, $\forall~s\in \disc{1}{n}$, $\forall~\bolds{l_u}\in \disc{1}{L}^{|u|}$. \\

%
Let us remark that it would be numerically very expensive to consider the estimation of all these coefficients. In practice, we make a truncation of the functional decomposition at a given order $d$, so that $Y\simeq \sum_{\substack{u\in S\\ |u|\leq d}} \eta_u(\mathbf{X_u})$. The relative size $d$ can be chosen by the notions of effective dimension~\cite{wang}\\ Even for small $d \ll p$, the number of terms blows up with the dimensionality of the problem, and so would the number of model evaluations when using an ordinary least-squares regression scheme. 
As an illustration, take $d=3$, $p=8$ and $L=5$. In this case, $m=7740$ parameters have to be estimated, which could be a difficult task in practice. \\
To handle this kind of problem, 
many variable selection methods have been considered in the field of statistics. The next section aims at briefly exposing the variable selection methods based on a penalized regression. We particularly focus on the $\ell_0$ penalty~\cite{temlyakov} and on the Lasso regression~\cite{tibshirani}.


\subsection{The variable selection methods}\label{sect42}

For simplicity, we denote by $m$ the number of parameters in (\ref{minmatrix}). The variable selection methods usually deal with the penalized regression problem
\begin{equation}\label{penalisation}
\min_{\boldsymbol\beta}  \| \mathbb Y-\mathbb X_{\hat \phi} \boldsymbol\beta\|_{\text{\tiny n}}^2 + \lambda J(\boldsymbol\beta)
\end{equation}

where $J(\cdot)$ is positive valued for $\boldsymbol\beta\neq 0$, and where $\lambda \geq 0$ is a tuning parameter. The most intuitive approach is to consider the $\ell_0$-penalty $J(\boldsymbol\beta)=\norm{\boldsymbol\beta}_0$, where $\norm{\boldsymbol\beta}_0= \sum_{j=1}^m \mathds 1 (\beta_j \neq 0)$. Indeed, the $\ell_0$ regularization aims at selecting non-zero coefficients, thus at removing the useless parameters from the model. 
The greedy approximation~\cite{temlyakov} offers a series of strategies to deal with the $\ell_0$-penalty.
Nevertheless, the $\ell_0$ regularization is a non convex function, and suffers from  statistical instability, as mentioned in~\cite{breiman,tibshirani}. The Lasso regression could be regarded as a convex relaxation of the optimization problem~\cite{tibshirani}.
Indeed, the Lasso regression is based on $\ell_1$-penalty, i.e. (\ref{penalisation}) with $J(\boldsymbol\beta)=\norm{\boldsymbol\beta}_1$, and $\norm{\boldsymbol\beta}_1=\sum_{j=1}^m \abs{\beta_j}$. The Lasso offers a good compromise between a rough selection of non-zero elements, and a ridge regression ($J(\boldsymbol\beta)=\sum_{j=1}^m \beta_j^2$) that only shrinks coefficients, but is known to be stable~\cite{buhlmann,hastie}. 
In the following, the proposed method will use either the $\ell_0$ or the $\ell_1$ regularization.\\
The adaptive forward-backward greedy (\emph{FoBa}) algorithm proposed in Zhang \cite{zhang} is exploited here to deal with the $\ell_0$ penalization. 
From a dictionary $\mathcal D$ that can be large and/or redundant, the \emph{FoBa} algorithm is an iterative scheme that sequentially selects and discards the element of $\mathcal D$ that has the least impact on the fit. The aim of the algorithm is to efficiently select a limited number of predictors. The advantage of such an approach is that it is very intuitive, and easy to implement. In our problem, the \emph{FoBa} algorithm is applied on the whole set of basis functions. It may happen that no basis function is retained for the estimation of a HOFD component. In this case, as we want to estimate each component of the HOFD, the corresponding coefficient is set to zero.\\
Initiated by Efron \textit{et al.}~\cite{efron}, the modified LARS algorithm is adopted to deal with the Lasso regression.  
The LARS is a general iterative technique that builds up the regression function by successive steps.
The adaptation of LARS to Lasso (the modified LARS) is inspired by the homotopy method proposed by Osborne \textit{et al.}~\cite{osborne}. The main advantage of the modified LARS algorithm is that it builds up the whole regularized solution path $\{\hat{\boldsymbol\beta}(\lambda), \lambda \in \mathbb{R}\}$, exploiting the property of piecewise linearity of the solutions with respect to $\lambda$~\cite{buhlmann,osborne2}.

In the next part, both the \emph{FoBa} and the modified LARS algorithms are adapted to our problem and are then compared on numerical examples. 


\subsection{Summary of the estimation procedure}\label{sectsum}

Provided an initial choice of orthonormal system $(\psi_{l_i})_{l_i=0,i\in \disc{1}{p}}^{L}$, we first construct the approximation spaces $\hat H_{u}^{0,L}$ of $H_u^0$ for $|u|\leq d$, and $d\ll p$, using the HOGS Procedure described in Section \ref{sect3}. A HOFD component $\eta_u$ is then a projection onto $\hat H_{u}^{0,L}$, whose coefficients are defined by least-squares estimation.
To bypass the curse of dimensionality, the \emph{FoBa} algorithm or the modified LARS algorithm is used. Once the HOFD components are estimated, we derive the empirical estimation of the generalized Sobol sensitivity indices given in Definition \ref{def1}.


\subsection{Asymptotic results}\label{sectasymp}

The method exposed in Section \ref{sect3} aims at estimating the ANOVA components, whose uniqueness is ensured by hierarchical orthogonality. However, we would like to make sure that our numerical procedure is robust, i.e. that the estimated summands converge to the theoretical ones. To do that, we 
suppose that the model function $\eta$ is approximated by $\eta^R$, where
\[
 \eta^R(\mathbf X)=\sum_{u \in S} \eta_u^R (\mathbf X), \quad \text{with}~~  \eta_u^R=\sum_{\bolds{l_u}\in \disc{1}{L}^{|u|}} \beta_{\bolds{l_u}}^{u,0} \phi_{\bolds{l_u}}(\mathbf{X_u}).
\]

In addition, we assume that the dimensionality of the problem under consideration is small,
i.e. that we obtain the estimator  $\hat\eta^R$ of $\eta^R$ given
\[ 
 \hat\eta^R(\mathbf X):=\sum_{u \in S}\hat\eta_u^R(\mathbf{X_u}), \quad \text{with}~~  \hat\eta_u^R(\mathbf{X_u})=\sum_{\bolds{l_u}\in\disc{1}{L}^{|u|}} \hat\beta_{\bolds{l_u}} \hat \phi_{\bolds{l_u}}(\mathbf{X_u}),
 \]
 where the $(\hat\beta_{\bolds{l_u}}^u)_{\bolds{l_u}\in \disc{1}{L}^{|u|}}$, $u\in S$ are estimated by the usual least squares estimation (\ref{minmatrix}), and $ \hat \phi_{\bolds{l_u}}\in \hat H_u^{0,L}$. Here, we are interested in the consistency of $\hat \eta^R$ when the dimension $L^{|u|}$ is fixed for all $u\in S$. This result is stated in Proposition \ref{procv}. 

\begin{proposition}\label{procv}
Assume that
 \begin{equation*}
Y=\eta^R(\mathbf X)+\varepsilon, \quad \textrm{where}~~ \eta^R(\mathbf X)=\sum_{u\in S}  \sum_{\bolds{l_u}\in\disc{1}{L}^{|u|}}  \bolds\beta^{u,0}_{\bolds{l_u}} \phi_{\bolds{l_u}}(\mathbf{X_u}),
\end{equation*}
with ~$\mathbb E(\varepsilon)=0$, and $\mathbb{E}(\varepsilon^2)=\sigma_*^2$, ~$\mathbb E(\varepsilon\cdot\phi_{\bolds{l_u}}(\mathbf{X_u}))=0$, $\forall~\bolds{l_u}\in\disc{1}{L}^{|u|}$, $\forall~u \in S$. ($\bolds\beta_0=(\bolds\beta^{u,0}_{\bolds{l_u}})_{\bolds{l_u},u}$ is the true parameter).\\

  Further, let us consider the least squares estimation $\hat\eta^R$ of $\eta^R$ using the sample $(y^s,\mathbf x^s)_{s\in\disc 1n}$ from the distribution of $(Y,\mathbf X)$, and the functions $({\hat\phi}_{\bolds{l_u}})_{\bolds{l_u}}$, that is
 \[
 \hat\eta^R(\mathbf X)=\sum_{u \in S}\hat\eta_u^R(\mathbf{X_u}), \quad \textrm{where}~~ \hat\eta_u^R(\mathbf{X_u})=\sum_{\bolds{l_u}\in \disc{1}{L}^{|u|}} \hat\beta_{\bolds{l_u}}^u \hat \phi_{\bolds{l_u}}(\mathbf{X_u}),
 \] 
 where $\hat{\bolds{\beta}}=\argmin_{\bolds\beta\in \Theta} \normemp{\mathbb Y-\mathbb X_{\hat \phi} \bolds\beta}^2$.\\


 If we assume that 
 \renewcommand{\labelenumi}{(H.\arabic{enumi})} 
 \begin{enumerate}
\item \label{hypcv1} The distribution $P_{\mathbf X}$ is equivalent to $\otimes_{i=1}^p P_{X_i}$;
%
\item \label{hypcv5} For any $u \in S$, any $\bolds{l_u}\in \disc 1{L}^{|u|}$, $\norm{\phi_{\bolds{l_u}}}=1$ and $\normemp{\hat \phi_{\bolds{l_u}}}=1$
\item \label{hypcv4} For any $i\in \disc 1 p$, any $l_i\in \disc 1 {L}$, $\norm{\hat \phi_{l_i}^2}<+\infty$.
\end{enumerate}

Then, 
\begin{equation}\label{convergence}
\alsur{\norm{\hat\eta^R-\eta^R}} 0~~ \textrm{when} ~n\rightarrow +\infty.
\end{equation}
\end{proposition}

The proof of Proposition \ref{procv} is postponed to Appendix \ref{appendA}. \\

Our aim here is to study how the approximating spaces $\hat H_{u}^{0,L}$, constructed with the previous procedure, behave when $n \rightarrow +\infty$. However, we assume that 
the dimension of $H_u^{0,L}$, $L^{|u|}$ is fixed. By extending the work of Stone~\cite{stone}, Huang~\cite{huang} explores the convergence properties of functional ANOVA models when $L^{|u|}$ is not fixed anymore. Nevertheless, the results are obtained for a general model space $H_u$, and its approximating space $\hat H_u$, $\forall~u\in S$. In~\cite{huang}, the author states that if the basis functions are $m$-smooth and bounded, $\norm{\hat\eta-\eta}$ converges in probability. For polynomials, Fourier transforms or splines, he specifically shows that $\norm{\hat\eta-\eta}=O_p(n^{-\frac{2m}{2m+d}})$ (See~\cite{huang} p. 257), when $d$ is the ANOVA order. Thus, even if we show the convergence of $\hat \eta^R$ for $d=p$, where $p$ is the model dimension, it is in our interest to have a small order $d$ when $p$ gets large to get a good rate of convergence. Also, it should be noticed that estimating the whole set of interaction terms is infeasible in practice when the model dimension is above $p=4$. For these reasons, we substitute the initial model for a functional ANOVA of order at most $d=2$ in the next numerical applications.


\section{Application}\label{sectappli}

In this section, we are interested by the numerical efficiency of the HOGS procedure introduced in Section \ref{sect3}, that may be coupled with a penalized regression, as done in Section \ref{sect4}. The goal of the following study is to show that our strategy gives a good estimation of the generalized sensitivity indices in an efficient way.
\subsection{Description}
In this study, we compare several numerical strategies summarized further below.
\begin{enumerate}
\item The HOGS Procedure consists in constructing the basis functions that will be used to estimate the components of the functional ANOVA. Further, to estimate the $m$ unknown coefficients, we may use
\begin{enumerate}
\item the usual least squares estimation when $m<n$, and $n$ is the number of model evaluations. This technique is called LSEHOGS.
\item when $m\geq n$, the HOGS can be coupled with the adaptive greedy algorithm FoBa to solve the $\ell_0$-penalized problem. This is called FoBaHOGS. To relax the $\ell_0$ penalization, the modified LARS algorithm may replace the greedy strategy, abbreviated LHOGS, where L stands for LARS.
\end{enumerate}
\item The general method developed in~\cite{chastaing}, based on projection operators, consists in solving a functional linear system, when the model depends on independent pairs of dependent inputs. This procedure is abbreviated POM for Projection Operators Method.
\item At last, we compare our strategy to a minimization under constraints detailed in~\cite{hooker}, and summarized as
\begin{equation}\label{optimatrix0}
 \left\{
\begin{array}{l}
  \min_{F} \normemp{\mathbb Y-\mathbb XF}^2\\
D_n F=0
\end{array}
\right.
\end{equation}
with $\mathbb Y_s=y^s-\bar y$, $s=1,\cdots,n$, $\mathbb X$ is a matrix composed of $1$ and $0$ elements, $F_{s,u}=\eta_u(\mathbf{x_u}^s)$, and $D_n$ is the matrix of hierarchical orthogonal constraints, where the inner product $\psth {\cdot, \cdot}$ has been replaced with its empirical version $\psemp \cdot \cdot$.
However, the matrix $\mathbb X$ is not full rank, so the solution of (\ref{optimatrix0}) is not unique. This implies that the descent technique used to estimate $F$ may give local solutions and lead to wrong results. Moreover, in a high-dimensional paradigm, the matrix of constraints $D_n$ becomes very large, leading to an intractable strategy in practice. To remedy to these issues, we consider that each component is parametrized, and constrained, i.e. we consider the following problem,
\begin{equation}\label{optimatrix}
 \left\{
\begin{array}{l}
  \min_{\Phi} \normemp{\mathbb Y-\mathbb X\Phi}^2\\
D_n\Phi=0
\end{array}
\right.
\end{equation}
where $\mathbb X_{s,\bolds{l_u}}=\psi_{\bolds{l_u}(\mathbf {x_u}^s)}$, $s=1,\cdots,n$, with $(\psi_{\bolds{l_u}})_{\bolds{l_u},u}$ the usual tensor basis of $L^2(\mathbb R)$ (polynomial, splines, \ldots). The vector $\Phi$ is the set of unknown coefficients, and $D_n$ is the matrix of constraints given by (\ref{hortho}), where the inner product $\psth {\cdot, \cdot}$ has been replaced with its empirical version $\psemp \cdot \cdot$, on the parametrized functionals of the decomposition. The Lagrange function associated with (\ref{optimatrix}) can be easily derived, and the linear system to be solved is the following
\begin{equation*}\label{syst}
 \begin{pmatrix}
  {}^t \mathbb X\mathbb X & -{}^t D \\
D & 0
 \end{pmatrix}
\begin{pmatrix}
 \Phi \\
\lambda
\end{pmatrix}=
\begin{pmatrix}
 {}^t \mathbb X\mathbb Y\\
0
\end{pmatrix}, 
\end{equation*}
where $\lambda$ is the Lagrange multiplier. This procedure, substantially similar to the HOGS Procedure, is abbreviated MUC for Minimization Under Constraints.
\end{enumerate}

In the following, all these strategies are compared in terms of computational time, as well as mean squared error, defined as
$$
\textrm{mse}(\eta)=\frac 1 {|S|} \sum_{u \in S} \frac 1 n \sum_{s=1}^n \left[\hat \eta_u(\mathbf {x_u}^s)- \eta_u(\mathbf{x_u}^s)\right]^2,
$$
 where the functions $\hat \eta_u$ are estimated by one of the methodologies described above, and $\eta_u$ are the analytical functions.
 \subsection{Test cases and results}
For every model, we consider that the functional ANOVA decomposition is truncated at order $d=2$. Also, we choose polynomial basis to make a proper comparison with the MUC strategy.
\paragraph*{\textbf{Test case 1: the g-Sobol function.}} Well known in the sensitivity literature~\cite{saltelli}, the g-Sobol function is given by
\[
Y=\prod_{i=1}^p \frac{\abs{4X_i-2}+a_i}{1+a_i}, \quad a_i \geq 0,
\]
where the inputs $X_i$ are independent and uniformly distributed over $[0,1]$. The analytical Sobol indices are given by
\[
S_u=\frac{1}{D}\prod_{i\in u} D_i, \quad D_i=\frac{1}{3(1+a_i)^2},~~D=\prod_{i=1}^p (D_i+1)-1,\quad \forall~u\in S.
\]
We take $p=10$, and $a=(0,1,4.5,9,99,99,99,99,99,99)$. We choose the Legendre polynomial basis of degree $5$, and we make $n=1500$ model evaluations over $50$ runs. Therefore the number of parameters to be evaluated is $m=1175<n$. The purpose of this study is to show the efficiency of the HOGS strategy, and to compare it with the MUC one. Our aim is also to consolidate the asymptotic result given in Section \ref{sect4} from a numerical viewpoint. Figure \ref{gsobol} plots the estimated sensitivity indices when LSEHOGS and MUC methods are used. The computational efficiency of both strategies are reported in Table \ref{tablegsobol}.
\begin{figure}[h]
\begin{center}
\subfigure[][LSEHOGS]{
\resizebox*{10cm}{!}{\includegraphics[angle=-90]{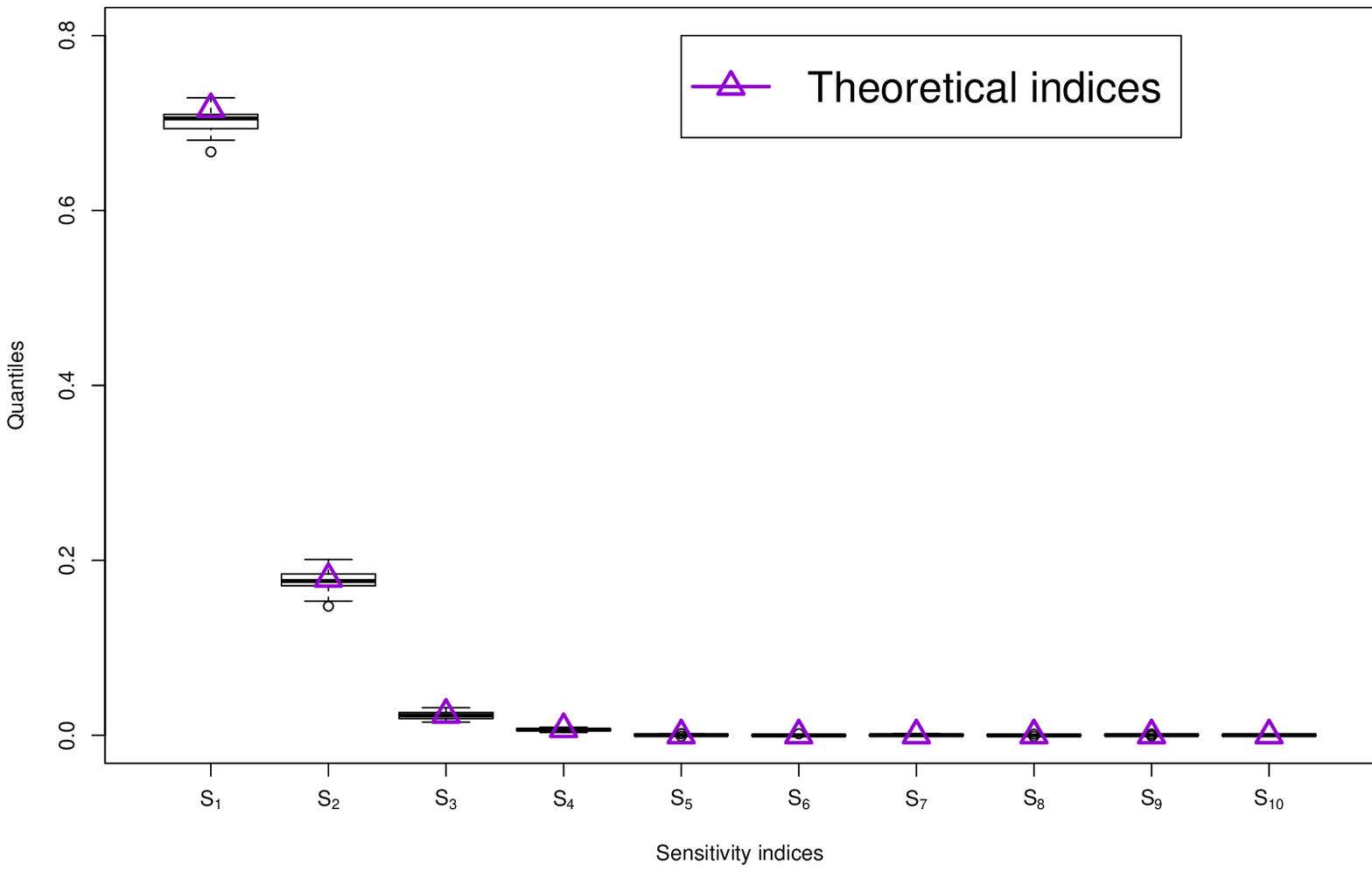}}}\\
\subfigure[][MUC]{
\resizebox*{10cm}{!}{\includegraphics[angle=-90]{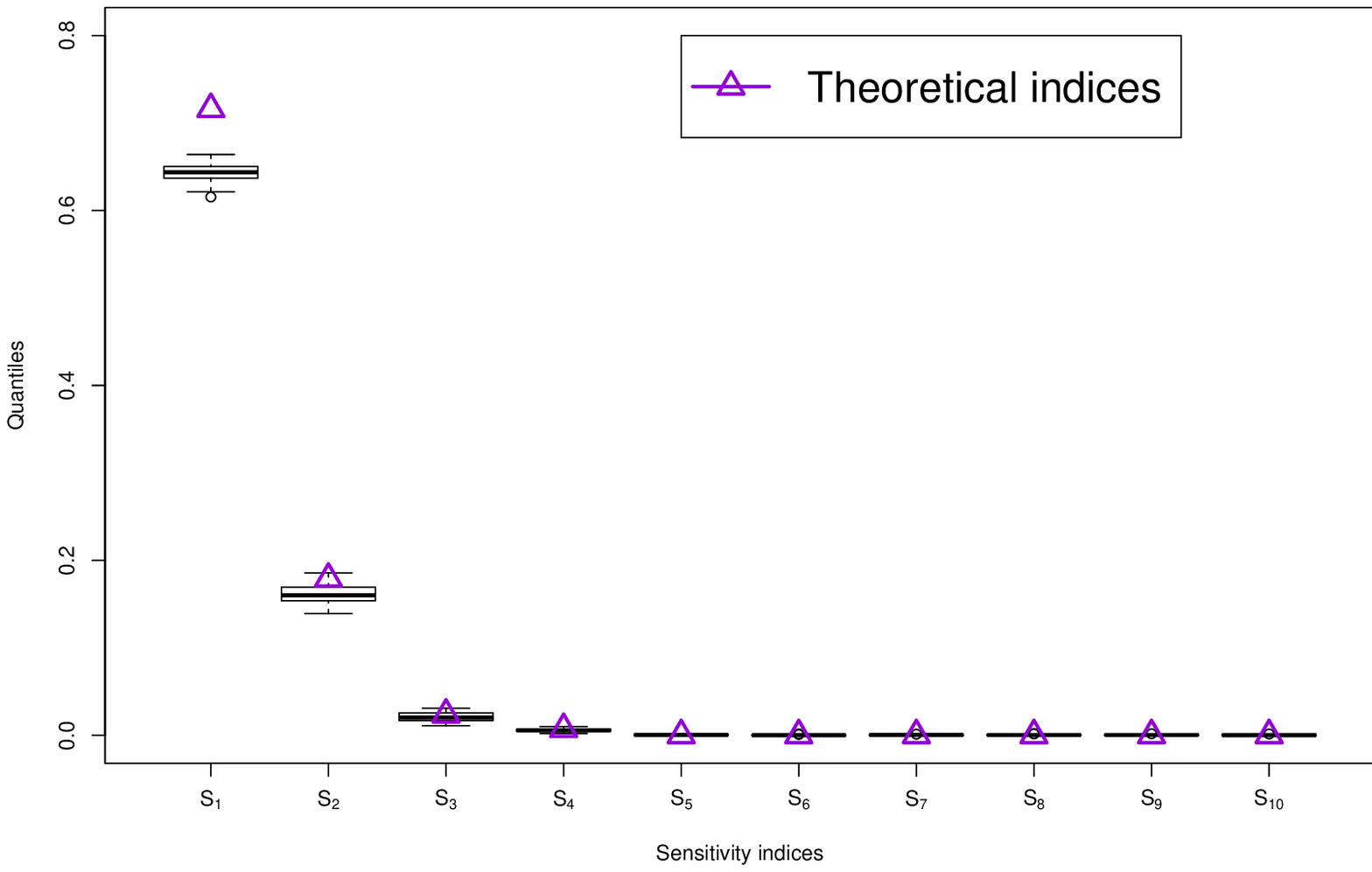}}}
\caption{\label{gsobol} Test case 1. Sensitivity indices estimation }
\label{figcopula}
\end{center}
\end{figure}
\begin{table}[!h]
\tbl{Test case 1. Numerical comparisons of the MUC and the LSEHOGS}
{\begin{tabular}{ccc}
\toprule
 & CPU time (in sec.) & $\textrm{mse}(\eta)\times 10^{-3}$ \\
\colrule
MUC & 64.02 &  3.58 \\
\colrule
LSEHOGS & 19.65 & 1.18 \\
\botrule
\end{tabular}}\label{tablegsobol}
\end{table}
\paragraph*{\textbf{Test case 2: the Li function.}} This polynomial function has been introduced by Li \textit{et al.}~\cite{li2}, and is given by the following expression,
\[
Y=g_1(X_1,X_2)+g_2(X_2)+g_3(X_3), \quad \mathbf X\sim N(0,\Sigma),
\]
with
\begin{eqnarray*}
 g_1(X_1,X_2)&=&[a_1X_1+a_0][b_1X_2+b_0]\nonumber\\
g_2(X_2)&=&c_2X_2^2+c_1X_2+c_0\nonumber\\
g_3(X_3)&=&d_3X_3^3+d_2X_3^2+d_1X_3+d_0\nonumber\\
 \Sigma&=&
\begin{pmatrix}
 \sigma_1^2 & \gamma\sigma_1\sigma_2 & 0\\
\gamma\sigma_1\sigma_2 &\sigma_2^2 & 0\\
0& 0 &  \sigma_3^2 
\end{pmatrix}. \nonumber
\end{eqnarray*}

The normal distribution does not satisfy Condition (\ref{c2}). 
However, the Gaussian density makes it possible to compute a HOFD decomposition, as done in~\cite{li2}. Moreover, if the research of solutions is restricted to the polynomial spaces, the uniqueness of the HOFD components given in~\cite{li2} is ensured, whatever the type of distribution. Thus, the analytical form of the ANOVA components and the generalized Sobol indices can be derived in this case. \\
To mimic the work done in~\cite{li2}, we take $a_0=c_1=d_0=1$, $a_1=b_0=c_2=d_1=d_2=2$ and $b_1=c_0=d_3=3$. The variations are set equal to $\sigma_1=\sigma_2=0.2$, $\sigma_3=0.18$ and $\gamma=0.6$. 
 For each component, we choose a Hermite basis of degree $10$ to apply the HOGS Procedure and the MUC strategy. Thus, the number of parameters to be estimated is equal to $m=330$. Further, we consider $n=300$ model evaluations to estimate the parameters by the (L/FoBa)HOGS method. We repeat the test over $50$ repetitions. We compare it to the MUC and the POM strategies in Table \ref{tablegen2} on the estimated sensitivity indices. Table \ref{tablegen3} shows the number of non-zero estimated coefficients for the FoBaHOGS and the LHOGS. The averaged elapsed time and the $\mathrm{mse}(\eta)$ computed for each method are also reported in Table \ref{tablegen3}.

\begin{table}[!h]
\tbl{Test case 2. Sensitivity indices estimation (with standard deviations)}
{\begin{tabular}{cccccccc}
 \toprule
 &$S_1$ & $S_2$ & $S_3$ & $S_{12}$ & $S_{13}$ & $S_{23}$ & $S_{123}$  \\
\colrule
Analytical &  0.4683 &   0.4652  &  0.0593  &  0.0072   &      0    &     0  &       0  \\
\colrule
\multirow{2}{1.3cm}{POM} &   0.4402  &  0.4718 &   0.0810 &  -0.0014& -& -&-  \\
& (0.021) & (0.0401) & (0.0012) & (0.001) & -& - &  \\
\colrule
\multirow{2}{1.3cm}{FoBaHOGS} &   0.4488  &  0.4699  &  0.0714  &  0.0041&         0  &       0&  0.0058
 \\
&     (0.0216) &    (0.0190) &     (0.0233) &     (0.0028) &         (0) &          (0) &  \\
\colrule
\multirow{2}{1.3cm}{LHOGS} &  0.4536 &   0.4733  &  0.0745  &  0.0065  &  0.0013  & 0.0006& -0.0098
 \\     
&    (0.0216) &     (0.0193) &     (0.0227) &     (0.0017) &    (0.0017) &     (0.0009) &  \\       
\colrule
\multirow{2}{1.3cm}{MUC} & 0.4439  &  0.4533  &  0.0713  &  0.0002 &   0.0002  &  0.0001 & 0.0310 \\
& (0.0206) &    (0.0185) &     (0.0221) &    (0.0001) &    (0.0004) &   (0.0003) & \\
\botrule
\end{tabular}}\label{tablegen2}
\end{table}

 \begin{table}[!h]
\tbl{Test case 2. Numerical comparisons of the MUC, the POM and the FoBa/LHOGS}
{\begin{tabular}{cccc}
 \toprule
  & $\sum_j \mathds{1}(\hat\beta_j \neq 0)$ &   CPU time (in sec.) & $\textrm{mse}(\eta)\times 10^{-3}$\\
\colrule
POM  & -& 30.02  & 6.5\\
\colrule
MUC &  - & 30.89 & 7.9\\
\colrule
FoBaHOGS & 5.9& 11.83 & 5.1\\
\colrule
LHOGS&  152.6 & 168.06 & 5.1 \\     
\botrule
\end{tabular}}\label{tablegen3}
\end{table}

\begin{figure}
\centering
 \includegraphics[width=0.4\textwidth]{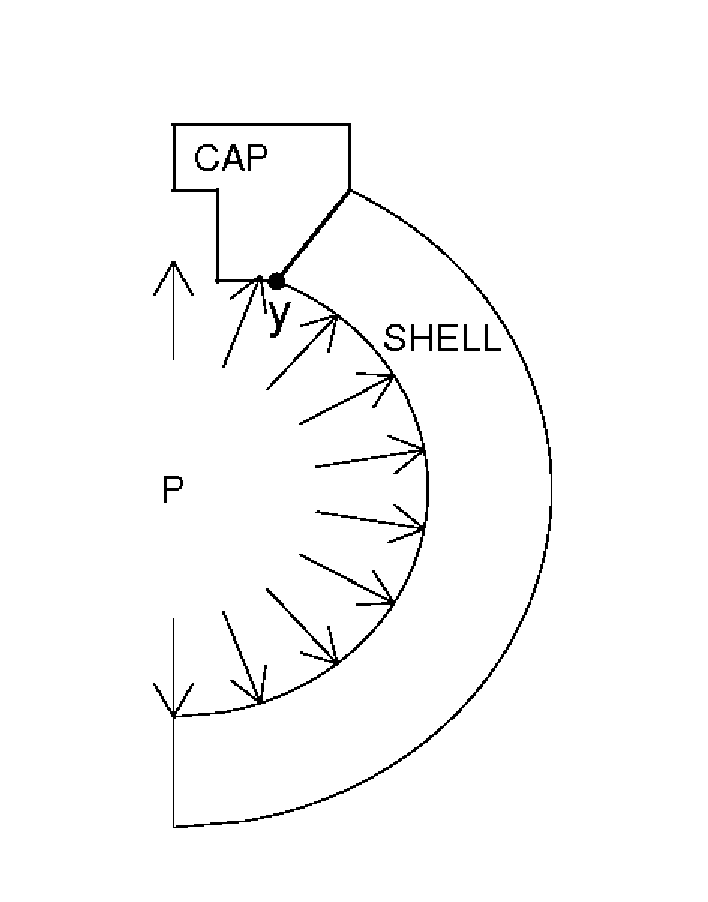}
\caption{Tank distortion at point $y$}\label{shell}
\end{figure}

\paragraph*{\textbf{Test case 3: the tank pressure model.}} The case study is a shell closed by a cap and subjected to an internal pressure. Figure \ref{shell} illustrates a simulation of tank distortion. The calculation of interest is the von Mises stress~\cite{vonmises} at the point $y$. The yielding of the material occurs as soon as the von Mises stress reaches the material yield strength.
The selected point $y$ corresponds to the point for which the von Mises stress is maximal in the tank. We want to prevent the tank from material damage induced by plastic strains. A 2D finite elements model the system using the code ASTER.
In order to design reliable structures, a manufacturer wants to identify the most contributive parameters to the von Mises criterion variability. 
The von Mises criterion depends on three geometrical parameters: $R_{int}$, $T_{shell}$ and $T_{cap}$. 
It also depends on six material parameters, $E_{shell}$, $E_{cap}$, $\sigma_{y,shell}$, $\sigma_{y,cap}$, and $P_{int}$. Table \ref{tabshell} gives the meaning and the distribution of the eight inputs.

\begin{figure}
\begin{center}
\subfigure[][$S_i$]{
\resizebox*{10cm}{!}{\includegraphics[angle=-90]{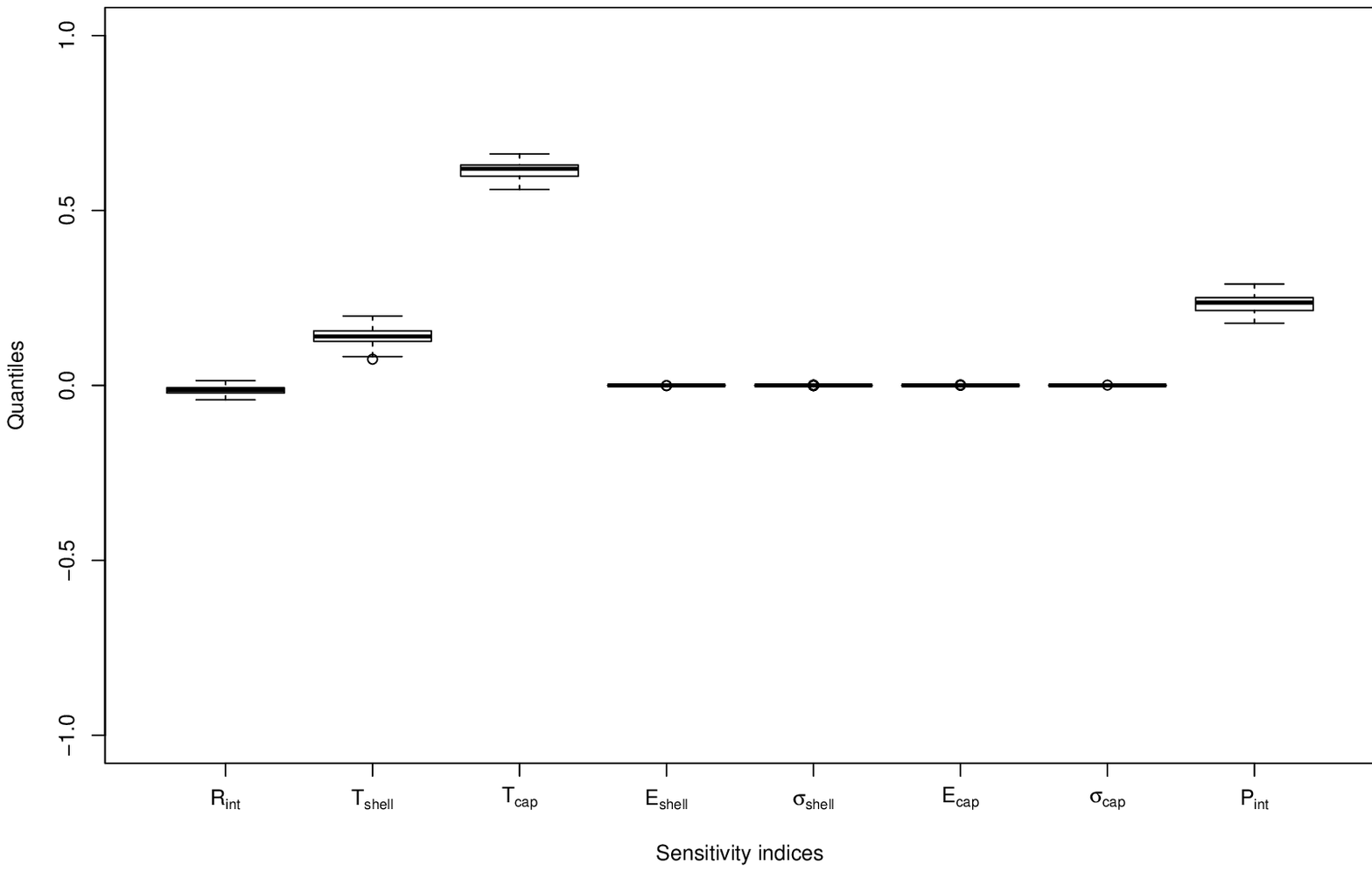}}}\\
\subfigure[][$\mathrm{VS}_i$]{
\resizebox*{6cm}{!}{\includegraphics[angle=-90]{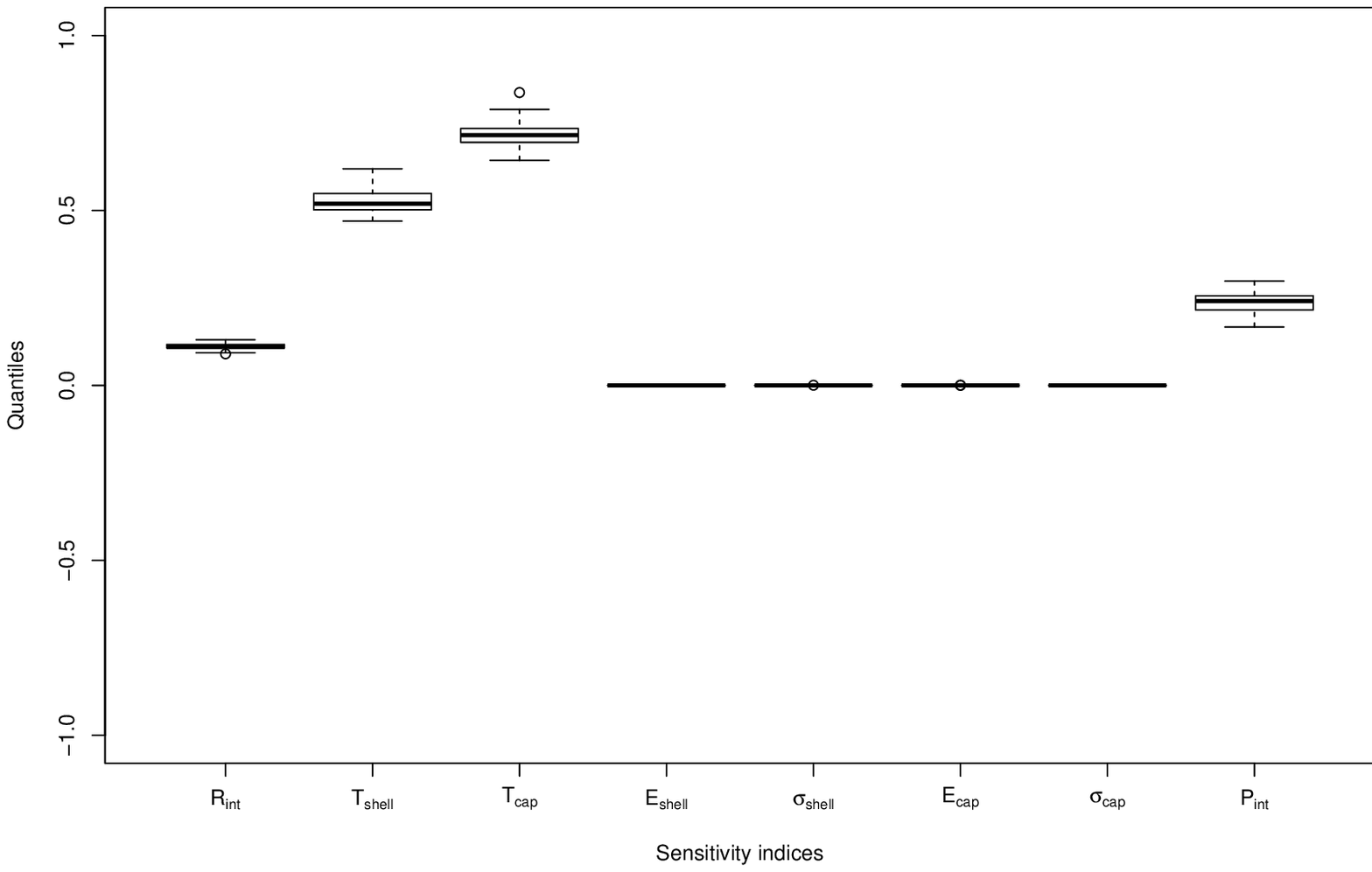}}\label{fig_vs}}\hspace{0.5cm}
\subfigure[][$\mathrm{CoVS}_i$]{
\resizebox*{6cm}{!}{\includegraphics[angle=-90]{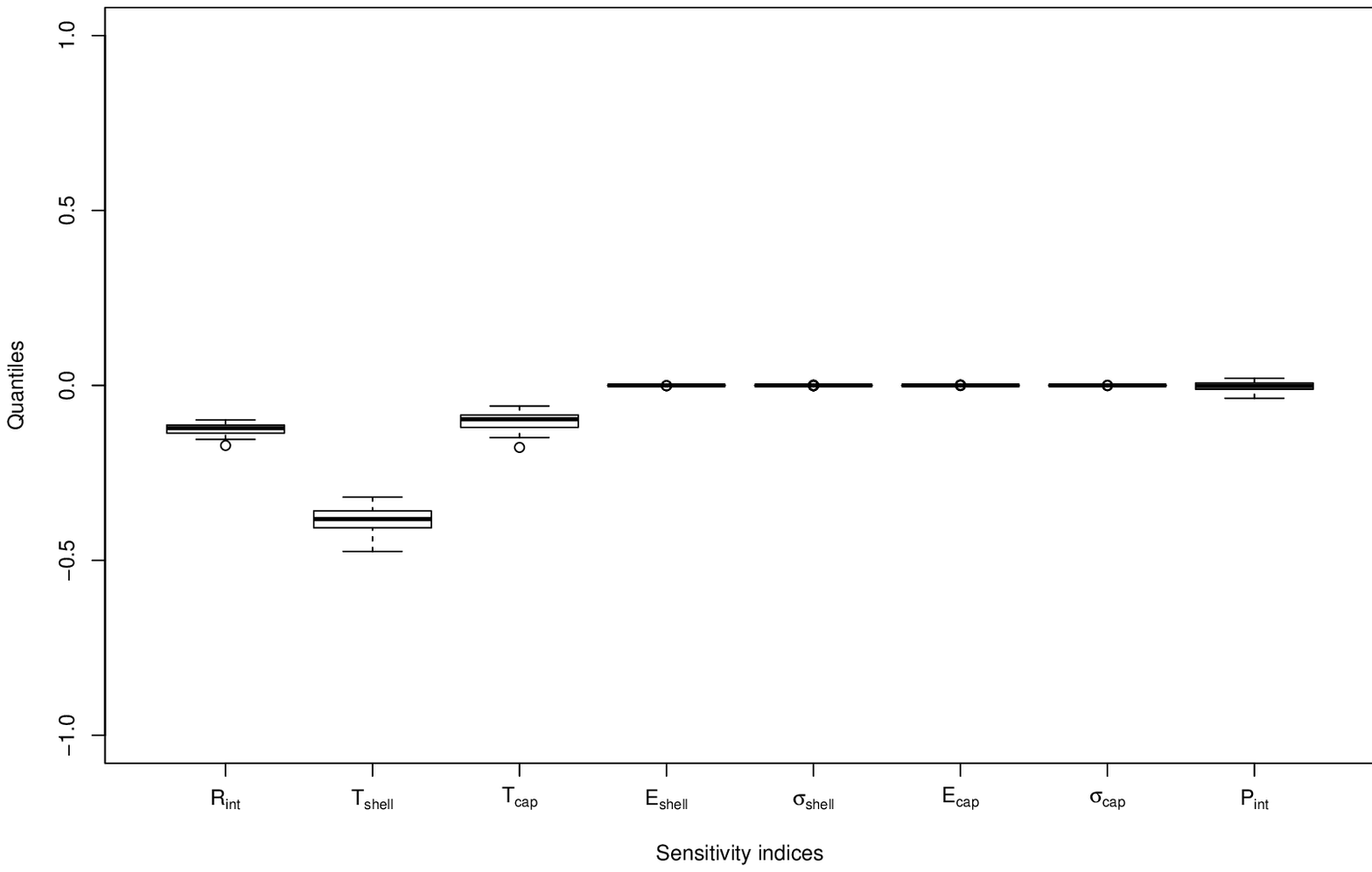}}\label{fig_covs}}
\caption{\label{gsobol} Test case 3. First order sensitivity indices}
\label{fig_vessel}
\end{center}
\end{figure}

\begin{table}
\tbl{Description of inputs of the shell model}
{\begin{tabular}{ccc}
\toprule
Inputs  & Meaning & Distribution\\
\colrule
$R_{int}$ & shell internal radius  &$\mathcal U([1800;2200])$, $\gamma(R_{int},T_{shell})=0.85$\\
$T_{shell}$  & shell thickness & $\mathcal U([360;440])$, $\gamma(T_{shell},T_{cap})=0.3$\\
$T_{cap}$  & cap thickness & $\mathcal U([180;220])$, $\gamma(T_{cap}, R_{int})=0.3$\\
\colrule
$E_{cap}$ & cap Young's modulus &  $\alpha N(\mu,\Sigma)+(1-\alpha)N(\mu,\Omega)$,  $\alpha=0.02$\\
$\sigma_{y,cap}$ & cap yield strength & $\mu=
\begin{pmatrix}
 210\\
500
\end{pmatrix}
$, $\Sigma=
\begin{pmatrix}
 350 & 0\\
0 &29
\end{pmatrix}
$, $\Omega=
\begin{pmatrix}
175 & 81 \\
81 & 417
\end{pmatrix}$   \\
\colrule
$E_{shell}$  & shell Young's modulus &$\alpha N(\mu,\Sigma)+(1-\alpha)N(\mu, \Omega)$, $\alpha=0.02$\\ 
$\sigma_{y,shell}$ & shell yield strength&  $\mu=
\begin{pmatrix}
 70\\
300
\end{pmatrix}
$, $\Sigma=
\begin{pmatrix}
 117 & 0\\
0 &500
\end{pmatrix}
$, $\Omega=
\begin{pmatrix}
58 & 37 \\
37 & 250
\end{pmatrix}$ \\
\colrule
$P_{int}$  & internal pressure & $N(80,10)$\\
\botrule
\end{tabular}}\label{tabshell}
\end{table}
 The geometrical parameters are uniformly distributed because of the large choice left for the tank building. The correlation $\gamma$ between the geometrical parameters is induced by the constraints of manufacturing processes. The physical inputs are normally distributed and their uncertainty are due to the manufacturing process and the properties of the elementary constituents variabilities. The large variability of $P_{int}$ in the model corresponds to the different internal pressure values which could be applied to the shell by the user.\\
To measure the contribution of the correlated inputs to the output variability, we estimate the generalized sensitivity indices by the FoBaHOGS method only, as LHOGS gives very similar results. 
We perform $n=1000$ simulations for each of $50$ replications. 
We use the $5$-spline functions for the geometrical parameters and the Hermite basis functions of degree $7$ for the physical parameters. 
Figure \ref{fig_vessel} displays the first order sensitivity indices $S_i$, $i=1,\cdots,8$, and their splits into $\mathrm{VS}_i$ and $\mathrm{CoVS}_i$ for an easier interpretation.

\paragraph*{\textbf{Interpretation.}}
It clearly appears in Case 1 that MUC underestimates some sensitivity indices, as observed in Figure \ref{gsobol}, whereas our strategy behaves well. Moreover, LSEHOGS is the fastest method, and gives a smaller mean squared error, although this quantity is low for both procedures.
When the model depends on dependent inputs, as in Case 2, we observe that the sensitivity indices computed with (L/FoBa)HOGS Procedure are close to the analytical ones, although every method leads to an estimation error. The indices obtained by MUC are slightly underestimated (especially for $S_{12}$), although their standard deviation is smaller than the one obtained for HOGS. 
Finally, POM demonstrates its limitations, as only interaction terms involved in dependent input pairs can be estimated. Finally, the (L/FoBa)HOGS outperforms all of its competitors, as the sensitivity indices are well estimated, even in a high-dimensional paradigm. Also, $\mathrm{mse}(\eta)$ computed by HOGS is  smaller than the one obtained by MUC and POM, and also smaller than the one provided by Li \textit{et al.}~\cite{li2}, where the authors use a continuous descent technique to estimate the components $(\eta_u)_{u\in S}$. The MUC method shows its limitations in a high-dimensional model, when the HOGS method has stable performance. In particular, the greedy strategy outperforms the other techniques in terms of computational cost. It also tends to estimate a very small number of non-zero coefficients in comparison with the Lasso, offering a numerical attractive strategy. However, it should be noted a larger variance for HOGS, compared to MUC. In view of the first two cases, it numerically appears that the HOGS Procedure, that may be coupled by a penalized regression, is unbiased, while MUC would be, and may lead to a larger mean-squared error.\\
The test case 3 is devoted to the sensitivity analysis. The material parameters  $E_{shell}$, $E_{cap}$, $\sigma_{y,shell}$, $\sigma_{y,cap}$ are independent from the other inputs, and their effects are negligible, so we can conclude that they do not have any influence in the model. The internal pressure $P_{int}$ has an influence on the model response, but 
the strongest contribution comes from the correlated set of geometrical inputs $(R_{int}, T_{shell}, T_{cap})$. Thanks to Figure \ref{fig_vs},\ref{fig_covs}, we deduce that $T_{cap}$ has an important full contribution, barely weakened by the contribution induced by the dependence. Thus, one can deduce that $T_{cap}$ is very significant in the model.  The sensitivity indices of $R_{int}$ and $T_{shell}$ are quite small, but it should be noticed that the covariance part plays the role of compensation, so one should reduce the variability of both parameters.

\section{Conclusions and perspectives}
This paper brings a new methodology to estimate the components of the generalized functional decomposition, when the latter satisfy hierarchical orthogonal constraints. Moreover, we show the consistency of the estimators when the usual least-squares estimation is used to estimate the unknown coefficients. 
From a practical point of view, it appears that the penalized regression should be often applied, and we observed that a greedy strategy is numerically efficient. The future objective is to go further into that point, i.e. to explore the numerical and theoretical properties of our methodology when the $\ell_0$ penalization is relaxed by the $\mathbb L_2$-boosting~\cite{friedman}. This approach is the object of a work in progress, inspired by the precursor work of~\cite{champion}.


\section*{Acknowledgments}

The authors would like to thank Loic Le Gratiet and Gilles Defaux for their help in providing and understanding the tank pressure model. \\
This work has been supported by French National Research Agency (ANR) through the COSINUS program (project COSTA-BRAVA
number ANR-09-COSI-015).


\newpage

\bibliographystyle{gSCS}
\bibliography{biblio_GS_Greedy(2)}


\appendices

\section{Convergence results}\label{appendA}

 In this part, we restate and prove Proposition \ref{procv} of Section \ref{sectasymp}. For sake of clarity, we first define and recall some notation that will be used further.
 
 \subsection*{Reminder}
 
 First, as mentioned in Section \ref{sect4}, we assume that $Y$ is centered. Recall that, $\forall~i\in \disc 1p$, $L$ is the dimension of the spaces $H_i^{0,L}$ and $\hat H_{i}^{0,L}$. 
Also, $\mbox{dim}(\h)=\mbox{dim}(\hath)=L^{|u|}$. For $u=\{u_1,\cdots,u_k\}\in S$, $\bolds{l_u}=(l_{u_1},\cdots,l_{u_k})$ is a multi-index of $\disc 1{L}^{|u|}$. \\

We refer $(\phi_{\bolds{l_u}})_{\bolds{l_u}\in \disc{1}{L}^{|u|}}$ as the basis of $H_u^{0,L}$ and  $(\hat \phi_{\bolds{l_u}})_{\bolds{l_u}\in \disc{1}{L}^{|u|}}$ as the basis of $\hat H_{u}^{0,L}$ constructed according to HOGS Procedure of Section \ref{sect3}. Thus, these functions all lie in $L^2_{\mathbb R}(\mathbb R^p,\mathcal{B}(\mathbb{R}^p),P_{\mathbf{X}})$. 

 $\psth{\cdot,\cdot}$ and $\norm\cdot$ are used as the inner product and norm on $L^2_{\mathbb R}(\mathbb R^p,\mathcal{B}(\mathbb{R}^p),P_{\mathbf{X}})$,
  \[
 \langle h_1,h_2\rangle=\int h_1(\mathbf x)h_2(\mathbf x) p_{\mathbf{X}} d\nu(\mathbf x), \quad \|h\|^2=\langle h,h\rangle, 
\]
 while $\psemp\cdot\cdot$ and $\normemp\cdot$ denote the empirical inner product and norm, that is
  \[
\langle g_1,g_2\rangle_{\textrm{\tiny n}}=\frac{1}{n}\sum_{s=1}^n g_1(\mathbf{x}^s)g_2(\mathbf{x}^s),\quad \|g\|_{\textrm{\tiny n}}^2=\langle g,g\rangle_{\textrm{\tiny n}},
\]

when $(y^s,\mathbf x^s)_{s=1,\cdots,n}$ is the $n$-sample of observations from the distribution of $(Y,\mathbf X)$.


\subsection* {New settings}

We set $m:=\sum_{u\in S} L^{|u|}$ the number of parameters in the regression model. Denote, for all $u\in S$, 
$
\Phi_u(\mathbf{X_u}) \in (L^2(\mathbb R, \mathcal B(\mathbb R), P_{\mathbf{X}}))^{L^{|u|}}$, with
$(\Phi_u(\mathbf{X_u}))_{\bolds{l_u}}=\phi_{\bolds{l_u}}(\mathbf{X_u})$, and by $\bolds\beta$ any vector of $\Theta\subset \mathbb R^{m}$, where $(\bolds\beta)_{\bolds{l_u},u}=\beta_{\bolds{l_u}}^u$, $\forall~\bolds{l_u}\in \disc{1}{L}^{|u|}$. \\

Recall that, for $a, b \in \mathbb N^*$, $\mathcal M_{a,b}(\mathbb R)$ denotes the set of all real matrices with $a$ rows and $b$ columns. \\

Set $\mathbb X_{\hat \phi}=\begin{pmatrix} \bolds{\hat \phi}_1 & \cdots & \bolds{\hat \phi}_u & \cdots\end{pmatrix} \in \cp{u\in S}\mathcal M_{n,L^{|u|}}(\mathbb R)$, where $(\bolds{\hat \phi}_u)_{s,\bolds{l_u}}=\hat \phi_{\bolds{l_u}}(\mathbf {x_u}^s)$,  and we set $\mathbb X_\phi=\begin{pmatrix}\bolds \phi_1 &\bolds \phi_2 & \cdots\end{pmatrix}\in \cp{u\in S}\mathcal M_{n,L^{|u|}}(\mathbb R)$, where $(\bolds \phi_u)_{s,\bolds{l_u}}=\phi_{\bolds{l_u}}(\mathbf {x_u}^s)$, for $u\in S$, $s\in \disc{1}{n}$ and $\bolds{l_u}\in\disc{1}L^{|u|}$. \\

Denote by $A_\phi$ be the $m \times m$ Gram matrix whose block entries are \linebreak $(\mathbb E(\Phi_u(\mathbf{X_u}) {}^t \Phi_v(\mathbf{X_v})))_{u,v\in S}$.\\


The main convergence result is reminded further below.

\begin{proposition}
Assume that
 \begin{equation*}
Y=\eta^R(\mathbf X)+\varepsilon, \quad \textrm{where}~~ \eta^R(\mathbf X)=\sum_{u\in S}  \sum_{\bolds{l_u}\disc{1}{L}^{|u|}}  \bolds\beta^{u,0}_{\bolds{l_u}} \phi_{\bolds{l_u}}(\mathbf{X_u}) \in H_u^{0,L},
\end{equation*}
with ~$\mathbb E(\varepsilon)=0$, $\mathbb{E}(\varepsilon^2)=\sigma_*^2$, ~$\mathbb E(\varepsilon\cdot\phi_{\bolds{l_u}}(\mathbf{X_u}))=0$, $\forall~\bolds{l_u}\in \disc{1}{L}^{|u|}$, $\forall~u \in S$. ($\bolds\beta_0=(\bolds\beta^{u,0}_{\bolds{l_u}})_{\bolds{l_u},u}$ is the true parameter).\\

  Further, let us consider the least-squares estimation $\hat\eta^R$ of $\eta^R$ using the sample $(y^s,\mathbf x^s)_{s\in\disc 1n}$ from the distribution of $(Y, \mathbf X)$, and the functions $(\hat \phi_{\bolds{l_u}})_{\bolds{l_u}}$, that is
 \[
 \hat\eta^R(\mathbf X)=\sum_{u \in S}\hat\eta_u^R(\mathbf{X_u}), \quad \textrm{where}~~ \hat\eta_u^R(\mathbf{X_u})=\sum_{\bolds{l_u}\in \disc{1}{L}^{|u|}} \hat\beta_{\bolds{l_u}}^u \hat \phi_{\bolds{l_u}}(\mathbf{X_u}) \in \hat H_{u}^{0,L},
 \] 
 where $\hat{\bolds{\beta}}=\argmin_{\bolds\beta\in \Theta} \normemp{\mathbb Y-\mathbb X_{\hat \phi}\bolds\beta}^2$. If we assume that 
 
 \renewcommand{\labelenumi}{(H.\arabic{enumi})} 
 \begin{enumerate}
\item \label{hypcv1bis} The distribution $P_{\mathbf X}$ is equivalent to $\otimes_{i=1}^p P_{X_i}$;
%
\item \label{hypcv5bis} For any $u \in S$, any $\bolds{l_u}\in \disc 1{L}^{|u|}$, $\norm{\phi_{\bolds{l_u}}}=1$ and $\normemp{\hat \phi_{\bolds{l_u}}}=1$
\item \label{hypcv4bis} For any $i\in \disc 1 p$, any $l\in \disc 1 {L}$, the fourth moment of $\hat \phi_{l_i}$ is finite.
\end{enumerate}
 
Then, 
\begin{equation}\label{convergence2}
\alsur{\norm{\hat\eta^R-\eta^R}} 0~~ \textrm{when} ~n\rightarrow +\infty.
\end{equation}
\end{proposition}


 The proof of Proposition is broken up into Lemmas \ref{noexist}-\ref{lemcv2}. To prove (\ref{convergence2}), we introduce $\bar\eta^R$ as the following approximation of $\eta^R$,
\[
\bar\eta^R=\sum_{u\in S}\bar\eta_u^R(\mathbf{X_u})= \sum_{u \in S}\sum_{\bolds{l_u}\in \disc{1}{L}^{|u|}} \beta_{\bolds{l_u}}^{u,0} \hat \phi_{\bolds{l_u}}(\mathbf{X_u})=\mathbb X_{\hat \phi} \bolds{\beta_0},
\]
and we write the triangular inequality,
 \begin{equation}\label{trineq}
 \|\hat\eta^R -\eta^R\|=\|\hat\eta^R -\bar\eta^R+\bar\eta^R-\eta^R\| \leq \|\hat\eta^R -\bar\eta^R\|+  \|\bar\eta^R -\eta^R\|.
 \end{equation}

\bigskip

Thus, it is enough to prove that $\alsur{\norm{\hat\eta^R-\bar\eta^R}}{0}$, and that $\alsur{\norm{\bar\eta^R-\eta^R}}{0}$.

Lemmas \ref{lemcv1} and \ref{lemcv2} deal with convergence results on $\|\bar\eta^R -\eta^R\|$ and on $\|\hat\eta^R -\bar\eta^R\|$, respectively.
Lemmas \ref{noexist}, \ref{res1} are preliminary results to prove Lemmas \ref{lemcv1} and \ref{lemcv2}.\\

 \subsection{Preliminary results}

 \begin{lemma}\label{noexist}
If (H.1) holds, then $A_\phi$ is a non singular matrix.
\end{lemma}
 
 
 \begin{proof}[{Proof of Lemma \ref{noexist}}]
 \mbox{}
 
First of all notice that when we consider a Gram matrix, 
by a classical argument on the associated quadratic form, the full rank of this matrix holds if and only if the associated functional vector has full rank in $L^2$~\cite{holmes}. \\

To begin with, set, for all $i\in\disc 1p$, $\Psi_i=\begin{pmatrix} 1 \\ \Phi_i \end{pmatrix}$ and $G_i:=\mathbb E(\Psi_i {}^t \Psi_i)$. As $(\phi_{l_i})_{l_i=1}^{L}$ is an orthonormal system, we obviously get $G_i=\id_{(L+1)\times (L+1)}$, where $\id$ denotes the identity matrix. Now we may rewrite the tensor product $\otimes_{i=1}^p G_i$ as
\begin{equation}\label{prodgi}
\otimes_{i=1}^p G_i=\otimes_{i=1}^p \mathbb E(\Psi_i {}^t \Psi_i)=\int\otimes_{i=1}^p\left[\Psi_i(x_i) {}^t \Psi_i(x_i)\right] dP_{X_1}(x_1)\cdots dP_{X_p}(x_p).
\end{equation}

We obviously have $\otimes_{i=1}^p G_i=\id$. So that, using the remark of the beginning of the proof, the system~ $\otimes_{i=1}^p\left[\Psi_i {}^t \Psi_i\right]=\begin{pmatrix}
1 & (\otimes_{i\in u}\Phi_i)_{\substack{u \subseteq \disc{1}{p}\\ u\neq \emptyset}} \end{pmatrix}$ is linearly independent $(\otimes_{i=1}^p P_{X_i})-$a.e. \\

As we assumed that~ $\otimes_{i=1}^p P_{X_i}$ and $P_{\mathbf{X}}$ are equivalent by (H.1), we get that $\begin{pmatrix}
1 & (\otimes_{i\in u}\Phi_i)_{\substack{u \subseteq \disc{1}{p}\\ u\neq \emptyset}} \end{pmatrix}$ is linearly independent $\pxae$.\\

Now, we may conclude as in the classical Gram-Schmidt construction. Indeed, the construction of the system $(\Phi_u)_{u\in S}$ involves an invertible triangular matrix.

 \end{proof}


\begin{lemma} \label{res1}
Let $u,v \in S$ and $\bolds{l_u}\in \disc{1}{L}^{|u|},~ \bolds {l_v}\in \disc{1}{L}^{|v|}$. Assume that (H.2) holds. Further, assume that $\alsur{\|\hat \phi_{\bolds{l_u}}-\phi_{\bolds{l_u}}\| }{0}$, $\alsur{\normemp{\hat \phi_{\bolds{l_u}}-\phi_{\bolds{l_u}}} }{0}$, $\alsur{\|\hat \phi_{\bolds{l_v}}-\phi_{\bolds{l_v }}\| }{0}$ and $\alsur{\normemp{\hat \phi_{\bolds{l_v}}-\phi_{\bolds{l_v }}} }{0}$.  Then, the following results hold:

\begin{enumerate}[(i)]
\item \label{itemi} $\alsur{\norm{\hat \phi_{\bolds{l_u} }}}{1}$ and $\alsur{\normemp{ \phi_{\bolds{l_u}}}}{1}$;
\item \label{itemii} $\alsur{\psth{\phi_{\bolds{l_u}},\hat \phi_{\bolds{l_v}}}}{\psth{\phi_{\bolds{l_u}},\phi_{\bolds{l_v}}}}$ and $\alsur{\psemp{\phi_{\bolds{l_u}}}{\hat \phi_{\bolds{l_v}}}}{\psth{\phi_{\bolds{l_u}},\phi_{\bolds{l_v}}}}$;
\item \label{itemiii} $\alsur{\psth{\hat \phi_{\bolds{l_u}},\hat \phi_{\bolds{l_v}}}}{\psth{\phi_{\bolds{l_u}},\phi_{\bolds{l_v}}}}$ and $\alsur{\psemp{\hat \phi_{\bolds{l_u}}}{\hat \phi_{\bolds{l_v}}}}{\psth{\phi_{\bolds{l_u}},\phi_{\bolds{l_v}}}}$;
\item \label{item2ii} For $u=\{u_1,\cdots,u_k\}\in S$, with $k\geq 1$, and $\bolds{l_u}\in\disc 1 {L}^{|u|}$, 
$$\alsur{\normemp{\prod_{i=1}^k \hat \phi_{l_{u_i}}- \prod_{i=1}^k \phi_{l_{u_i}}}} 0 ~~\implies ~~ \alsur{\psemp{\prod_{i=1}^k\hat \phi_{l_{u_i}}}{\hat \phi_{\bolds{l_v}}}}{\psth{\prod_{i=1}^k \phi_{l_{u_i}},\phi_{\bolds{l_v}}}}.$$
\end{enumerate}
\end{lemma}


\begin{proof}[{Proof of Lemma \ref{res1}}]
\mbox{}

The first point (\ref{itemi}) is trivial. Now, we have, by (H.2),

\beqe
\barr{lll}
\abs{\psth{\phi_{\bolds{l_u}},\hat \phi_{\bolds{l_v}}}-\psth{\phi_{\bolds{l_u}},\phi_{\bolds{l_v}}}} &=&\abs{\psth{\phi_{\bolds{l_u}},\hat \phi_{\bolds{l_v}}-\phi_{\bolds{l_v}}}}\\
&\leq & \underbrace{\norm{\phi_{\bolds{l_u}}}}_{=1} \alsur{\norm{\hat \phi_{\bolds{l_v}}-\phi_{\bolds{l_v}}}} 0.
\earr
\eeqe

Further,

\beqe
\barr{lll}
\abs{\psemp{\phi_{\bolds{l_u}}}{\hat \phi_{\bolds{l_v}}} - \psth{\phi_{\bolds{l_u}},\phi_{\bolds{l_v}}} } &\leq & \abs{\psemp{\phi_{\bolds{l_u}}}{\hat \phi_{\bolds{l_v}}-\phi_{\bolds{l_v}}} } + \abs{\psemp{\phi_{\bolds{l_u}}}{\phi_{\bolds{l_v}}} - \psth{\phi_{\bolds{l_u}},\phi_{\bolds{l_v}}} }\\
& = & \normemp{\phi_{\bolds{l_u}}}\normemp{\hat \phi_{\bolds{l_v}}-\phi_{\bolds{l_v}}}+\abs{\psemp{\phi_{\bolds{l_u}}}{\phi_{\bolds{l_v}}} - \psth{\phi_{\bolds{l_u}},\phi_{\bolds{l_v}}} }.
\earr
\eeqe

 By the usual strong law of large numbers, $\alsur{\abs{\psemp{\phi_{\bolds{l_u}}}{\phi_{\bolds{l_v}}} - \psth{\phi_{\bolds{l_u}},\phi_{\bolds{l_v}}} }} 0$. By hypothesis, $\alsur{\normemp{\hat \phi_{\bolds{l_v}}-\phi_{\bolds{l_v}}} } 0$, and $ \alsur{\normemp{\phi_{\bolds{l_u}}} }{ 1}$ by (\ref{itemi}). Hence, (\ref {itemii}) holds.\\

The point (\ref{itemiii}) follows from
$$ 
\begin{array}{lll}
\abs{\psth{\hat \phi_{\bolds{l_u}},\hat \phi_{\bolds{l_v}}}-\psth{\phi_{\bolds{l_u}},\phi_{\bolds{l_v}}}}&=& \abs{\psth{\hat \phi_{\bolds{l_u}}-\phi_{\bolds{l_u}},\hat \phi_{\bolds{l_v}}} + \psth{\phi_{\bolds{l_u}},\hat \phi_{\bolds{l_v}}-\phi_{\bolds{l_v}}}}\\
&\leq &  \norm{\hat \phi_{\bolds{l_u}}-\phi_{\bolds{l_u}}} \norm{\hat \phi_{\bolds{l_v}}}+ \norm{\phi_{\bolds{l_u}}}\norm{\hat \phi_{\bolds{l_v}}-\phi_{\bolds{l_v}}}.
\end{array}
$$
By assumptions, $\alsur{ \norm{\hat \phi_{\bolds{l_u}}-\phi_{\bolds{l_u}}}}{0}$ and $\alsur{\norm{\hat \phi_{\bolds{l_v}}-\phi_{\bolds{l_v}}}}{0}$. Thus, the first point of (\ref{itemiii}) is satisfied, as $ \alsur{\norm{\hat \phi_{\bolds{l_v}}}}{ \norm{\phi_{\bolds{l_v}}}}=1$ (by (\ref{itemi})). Further,

\beqe
\barr{lll}
\abs{\psemp{\hat \phi_{\bolds{l_u}}}{\hat \phi_{\bolds{l_v}}}-\psth{\phi_{\bolds{l_u}},\phi_{\bolds{l_v}}}} &\leq & \abs{\psemp{\hat \phi_{\bolds{l_u}}}{\hat \phi_{\bolds{l_v}}-\phi_{\bolds{l_v}}}}+\abs{\psemp{\hat \phi_{\bolds{l_u}}}{\phi_{\bolds{l_v}}}-\psth{\phi_{\bolds{l_u}},\phi_{\bolds{l_v}}}}\\
&=&\normemp{\hat \phi_{\bolds{l_u}}} \normemp{\hat \phi_{\bolds{l_v}}-\phi_{\bolds{l_v}}} 
 +\abs{\psemp{\hat \phi_{\bolds{l_u}}}{\phi_{\bolds{l_v}}}-\psth{\phi_{\bolds{l_u}},\phi_{\bolds{l_v}}}}.
\earr
\eeqe

First, $\normemp{\hat \phi_{\bolds{l_u}}}=1$. By hypothesis, $\alsur{\normemp{\hat \phi_{\bolds{l_v}}-\phi_{\bolds{l_v}}}} 0$. By (\ref{itemii}), $\alsur{\abs{\psemp{\hat \phi_{\bolds{l_u}}}{\phi_{\bolds{l_v}}}-\psth{\phi_{\bolds{l_u}},\phi_{\bolds{l_v}}}}} 0$, so we can conclude.\\

Let show (\ref{item2ii}). We have,

\beqe
\barr {lll}
\abs{\psemp{\prod_{i=1}^k\hat \phi_{l_{u_i}}}{\hat \phi_{\bolds{l_v}}}-\psth{\prod_{i=1}^k \phi_{l_{u_i}},\phi_{\bolds{l_v}}}} & \leq & \abs{\psemp{\prod_{i=1}^k\hat \phi_{l_{u_i}}-\prod_{i=1}^k \phi_{l_{u_i}}}{\hat \phi_{\bolds{l_v}}}}\\
&& + \abs{\psemp{\prod_{i=1}^k\phi_{l_{u_i}}}{\hat \phi_{\bolds{l_v}}} - \psth{\prod_{i=1}^k \phi_{l_{u_i}},\phi_{\bolds{l_v}}}}\\
&\leq & \normemp{\prod_{i=1}^k\hat \phi_{l_{u_i}}-\prod_{i=1}^k \phi_{l_{u_i}}} \\
&&+ \normemp{\prod_{i=1}^k\phi_{l_{u_i}}}\normemp{\hat \phi_{\bolds{l_v}}-\phi_{\bolds{l_v}}}\\

&&+ \abs{\psemp{\prod_{i=1}^k \phi_{l_{u_i}}}{\phi_{\bolds{l_v}}}- \psth{\prod_{i=1}^k \phi_{l_{u_i}},\phi_{\bolds{l_v}}}}.
\earr
\eeqe

By the strong law of large numbers,  $\alsur{\abs{\psemp{\prod_{i=1}^k \phi_{l_{u_i}}}{\phi_{\bolds{l_v}}}- \psth{\prod_{i=1}^k \phi_{l_{u_i}},\phi_{\bolds{l_v}}}}} 0$, and we can conclude with the previous arguments.

\end{proof}


\subsection{Main convergence results}


\begin{lemma}\label{lemcv1}

Remind that the true regression function is
\[
\eta^R(\mathbf X)=\sum_{u\in S}\eta_u^R(\mathbf{X_u}), ~\textrm{where}~~ \eta_u^R(\mathbf{X_u})=\sum_{\bolds{l_u}\in\disc{1}{L}^{|u|}} \beta_{\bolds{l_u}}^{u,0} \phi_{\bolds{l_u}}(\mathbf{X_u}).
\]
Further, let $\bar\eta^R$ be the approximation of $\eta^R$,
\[
\bar\eta^R(\mathbf X)=\sum_{u\in S}\bar\eta_u^R(\mathbf{X_u}), ~\textrm{where}~~ \bar\eta_u^R(\mathbf{X_u})=\sum_{\bolds{l_u}\in \disc{1}{L}^{|u|}} \beta_{\bolds{l_u}}^{u,0} \hat \phi_{\bolds{l_u}}(\mathbf{X_u}).
\]

Then, under (H.2)-(H.3), we have
 $$
 \alsur{\|\bar\eta_u^R -\eta_u^R\|}{0}\quad \forall~u \in S,\quad \textrm{and}~~ \alsur{\|\bar\eta^R -\eta^R\|}{0}.
 $$

\end{lemma}


\begin{proof}[{Proof of Lemma \ref{lemcv1}}]
\mbox{}

 For any $u\in S$, 
 
 $$\|\bar\eta_u^R -\eta_u^R\|=\| \sum_{\bolds{l_u}} \beta_{\bolds{l_u}}^{u,0} \hat \phi_{\bolds{l_u}}-\sum_{\bolds{l_u}} \beta_{\bolds{l_u}}^{u,0} \phi_{\bolds{l_u}}\| \leq \sum_{\bolds{l_u}} |\beta_{\bolds{l_u}}^{u,0}| \cdot\|\hat \phi_{\bolds{l_u}}-\phi_{\bolds{l_u}}\|.
 $$
 
  Let us show that $\alsur{\|\hat \phi_{\bolds{l_u}}-\phi_{\bolds{l_u}}\| }{0}$. Actually, the proof of this convergence requires the use of Lemma \ref{res1}, so we also have to show that $\alsur{\normemp{\hat \phi_{\bolds{l_u}}-\phi_{\bolds{l_u}}} }{0}$. These two results are going to be proved by a double induction on $|u|\geq 1$ and on $\bolds{l_u}\in \disc 1 {L}^{|u|}$. We set 
  
 \[
 (\mathcal H_k)\quad \forall~u, |u|=k, \quad 
 \left\{
 \barr{ll}
   \alsur{\|\hat \phi_{\bolds{l_u}}-\phi_{\bolds{l_u}}\| }{0} &\\
  \alsur{\normemp{\hat \phi_{\bolds{l_u}}-\phi_{\bolds{l_u}}} }{0} &\forall~ \bolds{l_u}\in \disc{1}{L}^{|u|}.
  \earr
  \right.
 \]
 Let us show that $(\mathcal H_k)$ is true for any $k\leq p$ :
\begin{itemize}
\item Let $u=\{i\}$, so $k=1$. We used the Gram-Schimdt procedure on $(\phi_{l_i})_{l_i=1}^{L}$ to construct $(\hat \phi_{l_i})_{l_i=1}^{L}$. Let us show by induction on $l_i$ that $\alsur{\|\hat \phi_{l_i} -\phi_{l_i}\|}{0}$, $\forall~s_i\in \disc{1}{L}$. Set
\[
(\mathcal H'_{l_i}) \quad 
\left\{
\barr{l}
\alsur{\|\hat \phi_{l_i}-\phi_{l_i}\|}{0}\\
\alsur{\normemp{\hat \phi_{l_i}-\phi_{l_i}}}{0}.
\earr
\right.
\]

\begin{itemize}
\item For $l_i=1$, $\hat \phi_{1}=\dfrac{\phi_{1}- \psemp{\phi_{1}}{\hat \phi_{0}}\hat \phi_{0}}{\underbrace{\normemp{\phi_{1}-\psemp{\phi_{1}}{\hat \phi_{0}}\hat \phi_{0}}}_{T_n^{1}}}$, with $\hat \phi_{0}=\phi_{0}=1$. So,

$$
\begin{array}{lll}
\|\hat \phi_{1}- \phi_{1}\| &\leq& \norm{\frac{1-T_n^{1}}{T_n^{1}}  \phi_{1}} +\norm{ \frac{ \psemp{\phi_{1}} {1} }{T_n^{1}}}.
\end{array}
$$
 As $(\phi_{1})_{l_i=1}^{L}$ is an orthonormal system, we get $\alsur{\abs{\psemp{\phi_{1}}{1}}} {\mathbb E(\phi_{1})}=0$ and $\alsur{T_n^{1}}{\norm{\phi_{1}}}=1$.
 
Therefore, $\alsur{\|\hat \phi_{1}- \phi_{1}\|}{0}$. Also,

$$
\begin{array}{lll}
\normemp{\hat \phi_{1}- \phi_{1}} &\leq& \frac{1-T_n^{1}}{T_n^{1}} \normemp{ \phi_{1}} + \frac{ \abs{\psemp{\phi_{1}} {1} }}{T_n^{1}}.
\end{array}
$$

Exactly with the same previous argument, we conclude that $\alsur{\normemp{\hat \phi_{1}- \phi_{1}}} 0 $, then $(\mathcal H'_1)$ is true.

\item Let $l_i\in \disc{1}{L}$. Suppose that $(\mathcal H'_k)$ is true for any $k \leq l_i$. Let us show $(\mathcal H'_{l_i+1})$ holds. By construction, we get,
\[
\hat \phi_{l_i+1}=\frac{\phi_{l_i+1}-\sum_{k=0}^{l_i} \langle \phi_{l_i+1},\hat \phi_{k}\rangle_{\textrm{\tiny n}}\cdot \hat \phi_{k}} {\underbrace{\|\phi_{l_i+1}-\sum_{k=0}^{l_i} \langle \phi_{l_i+1},\hat \phi_{k}\rangle_{\textrm{\tiny n}}\cdot \hat \phi_{k} \|_{\textrm{\tiny n}}}_{T_n^{l_i+1}}}.
\]
So,
\[
\norm{\hat \phi_{l_i+1}-\phi_{l_i+1}} \leq \norm{ \frac{1-T_n^{l_i+1}}{T_n^{l_i+1}}\phi_{l_i+1}}+\sum_{k=0}^{l_i} \norm{\frac{1}{T_n^{l_i+1}}\psemp{\phi_{l_i+1}}{\hat \phi_{k}}\cdot \hat \phi_{k}}.
\]

For all~$k\leq l_i$, 

$$
\begin{array}{lll}
\abs{\langle \phi_{l_i+1},\hat \phi_{k}\rangle_{\textrm{\tiny n}}}&=& \abs{ \psemp{\phi_{l_i+1}}{\phi_{k}}+\psemp{\phi_{l_i+1}}{\hat \phi_{k}-\phi_{k}}}\\
&\leq & \abs{ \psemp{\phi_{l_i+1}}{\phi_{k}}}+\abs{\psemp{\phi_{l_i+1}}{\hat \phi_{k}-\phi_{k}}}\\
&\leq & \abs{ \psemp{\phi_{l_i+1}}{\phi_{k}}}+\normemp{\phi_{l_i+1}}\cdot \normemp{\hat \phi_{k}-\phi_{k}}.
\end{array}
$$

By induction, $\alsur{\norm{\hat \phi_{k}-\phi_{k}}}{0}$. By the usual law of large numbers, $\alsur{\abs{ \psemp{\phi_{l_i+1}}{\phi_{k}}}}{\abs{ \psth{\phi_{l_i+1},\phi_{k}}}}=0$ as the system $(\phi_{l_i})_{l_i=1}^{L}$ is orthonormal. As $\alsur{\normemp{\phi_{l_i+1}}}{\norm{\phi_{l_i+1}}}=1$, we deduce that  $\alsur{\abs{\langle \phi_{l_i+1},\hat \phi_{k}\rangle_{\textrm{\tiny n}}}}{0}$. And,

\beqe
\norm{\langle \phi_{l_i+1},\hat \phi_{k}\rangle_{\textrm{\tiny n}} \hat \phi_{k} } \leq \norm{\langle \phi_{l_i+1},\hat \phi_{k}\rangle_{\textrm{\tiny n}} ^2}^{1/2} \underbrace{\norm{(\hat \phi_{k})^2}^{1/2}}_{<+\infty} \quad \textrm{by (H.3)}.
\eeqe
Also, 
$$
\begin{array}{lll}
\alsur{T_n^{l_i+1}}{\norm{\phi_{l_i+1}}}=1 \Rightarrow \alsur{\norm{\hat \phi_{l_i+1}-\phi_{l_i+1}}}{0}.
\end{array}
$$
Now,
\[
\normemp{\hat \phi_{l_i+1}-\phi_{l_i+1}} \leq  \frac{1-T_n^{l_i+1}}{T_n^{l_i+1}} \normemp{\phi_{l_i+1}}+\sum_{k=0}^{l_i} \frac{1}{T_n^{l_i+1}} \abs{\psemp{\phi_{l_i+1}}{\hat \phi_{k}}}.
\]
With the previous arguments, $\alsur{\psemp{\phi_{l_i+1}}{\hat \phi_{k}}} 0$, and $\alsur{T_n^{l_i+1}} 1$. Then, we conclude that $(\mathcal H'_{l_i+1})$ is true.

\end{itemize}

Therefore, $(\mathcal H_1)$ is satisfied. 

\item Let $k\in \disc{1}{p}$. Suppose now that $(\mathcal H_{|\tilde u|})$ is true for any $1\leq |\tilde u| \leq k-1$, and any $\bolds{l_{\tilde u}}\in \disc{1}{L}^{|u|}$. Show that $(\mathcal H_k)$ is satisfied. Let $u$ be such that $u=\{u_1,\cdots,u_k\}$. \\

First, as $(\mathcal H_{|\tilde u|})$ is true for any $1\leq |\tilde u| \leq k-1$, results (\ref{itemi})-(\ref{itemii})-(\ref{itemiii}) of Lemma \ref{res1} can be applied to any couple $(\phi_{\bolds{l_u}}, \hat \phi_{\bolds{l_u}})$ such that $|u|\leq k-1$.\\

Further, we have seen that, for any $\bolds{l_u}\in \disc 1{L}^{|u|}$,
\[
\hat \phi_{\bolds{l_u}}^{u}=\prod_{i=1}^k\hat \phi_{l_{u_i}}+\sum_{\substack{v \subset u\\ v \neq \emptyset}}\sum_{\bolds{l_v}\in \disc 1{L}^{|v|}} \lambda_{\bolds{l_v},\bolds{l_u}}^{n} \hat \phi_{\bolds{l_v}}+C^{n}_{\bolds{l_u}},
\]

where $(C^{n}_{\bolds{l_u}},( \lambda_{\bolds{l_v},\bolds{l_u}}^{n} )_{\bolds{l_v},v\subset u})$ are computed by the resolution of the following equations
\begin{equation}\label{condemp}
\left\{
\begin{array}{ll}
 \langle \hat \phi_{\bolds{l_u}},\hat \phi_{\bolds{l_v}}\rangle_{\textrm{\tiny n}}=0,& \forall v\subset u, ~\forall~\bolds{l_v} \in \disc{1}{L}^{|v|}\\
  \langle \hat \phi_{\bolds{l_u}},1\rangle_{\textrm{\tiny n}}=0.
\end{array}
\right.
\end{equation}

The resolution of (\ref{condemp}) leads to the resolution of a linear system, when removing $C^{n}_{\bolds{l_u}}$, of the type 
$$
A^{u,n} \Lambda^{u,n}=D^{\bolds{l_u},n},
$$ 

where $\Lambda^{u,n}$ is the vector of unknown parameters $(\lambda_{\bolds{l_v},\bolds{l_u}}^{n})_{\bolds{l_v}\in \disc{1}{L}^{|v|},v\subset u}$, $A^{u,n}$ is the matrix whose block entries are $(\psemp{\bolds{\hat \phi}_{v_1}}{\bolds{\hat \phi}_{v_2}})_{v_1,v_2\subset u}$, and $D^{\bolds{l_u},n}$ involves block entries $(-\psemp{\otimes_{i=1}^k \hat \phi_{l_{u_i}}}{\bolds {\hat \phi}_{v_i}})_{v_i\subset u}$. \\

Also, the theoretical construction of the functions $(\phi_{\bolds{l_u}})_{\bolds{l_u}}$ consists in setting

\[
\phi_{\bolds{l_u}}=\prod_{i=1}^k\phi_{l_{u_i}}+\sum_{\substack{v \subset u\\ v \neq \emptyset}}\sum_{\bolds{l_v}\in \disc 1{L}^{|v|}} \lambda_{\bolds{l_v},\bolds{l_u}} \phi_{\bolds{l_v}}+C_{\bolds{l_u}},
\]

where $(C_{\bolds{l_u}},( \lambda_{\bolds{l_v},\bolds{l_u}} )_{\bolds{l_v},v\subset u})$ are computed by the resolution of the following equations
\begin{equation}\label{condth}
\left\{
\begin{array}{ll}
 \langle \phi_{\bolds{l_u}},\phi_{\bolds{l_v}}\rangle=0,& \forall v\subset u, ~\forall~\bolds{l_v} \in \disc{1}{L}^{|v|}\\
  \langle \phi_{\bolds{l_u}},1\rangle=0.
\end{array}
\right.
\end{equation}

The resolution of (\ref{condth}) leads to the resolution of a linear system, when removing $C_{\bolds{l_u}}$, of the type 
$$
A^{u}_\phi \Lambda^{u}=D^{\bolds{l_u}},
$$ 
where $\Lambda^{u}$ is the vector of unknown parameters $(\lambda_{\bolds{l_v},\bolds{l_u}})_{\bolds{l_v}\in \disc{1}{L}^{|v|},v\subset u}$, $A^{u}_\phi$ is the matrix whose block entries are $(\psth{\bolds\phi_{v_1},\bolds\phi_{v_2}})_{v_1,v_2\subset u}$, and $D^{\bolds{l_u}}$ involves block entries $(-\psth{\otimes_{i=1}^k \phi_{l_{u_i}}, \bolds \phi_{v_i}})_{v_i\subset u}$. \\

We have,
\begin{equation}\label{systdegue}
\begin{array}{lll}
\norm{\hat \phi_{\bolds{l_u}}-\phi_{\bolds{l_u}}} &=& \|\prod_{i=1}^k\hat \phi_{l_{u_i}}-\prod_{i=1}^k\phi_{l_{u_i}} + \sum_{\substack{v \subset u\\ v \neq \emptyset}}\sum_{\bolds{l_v}\in  \disc 1{L}^{|v|}}( \lambda_{\bolds{l_v},\bolds{l_u}}^{n} \hat \phi_{\bolds{l_v}} -\lambda_{\bolds{l_v},\bolds{l_u}} \phi_{\bolds{l_v}})\\
&&+C^{n}_{\bolds{l_u}}-C_{\bolds{l_u}}\|\\
&\leq & \norm{\prod_{i=1}^k\hat \phi_{l_{u_i}}-\prod_{i=1}^k\phi_{l_{u_i}}} + \sum_{\substack{v \subset u\\ v \neq \emptyset}}\sum_{\bolds{l_v}\in \disc 1{L}^{|v|}}\norm{ \lambda_{\bolds{l_v},\bolds{l_u}}^{n} \hat \phi_{\bolds{l_v}} -\lambda_{\bolds{l_v},\bolds{l_u}} \phi_{\bolds{l_v}}}\\
&&+ \abs{C^{n}_{\bolds{l_u}}-C_{\bolds{l_u}}},
\end{array}
\end{equation}
and,
\begin{equation}\label{systdeguebis}
\begin{array}{lll}
\normemp{\hat \phi_{\bolds{l_u}}-\phi_{\bolds{l_u}}} 
&\leq & \normemp{\prod_{i=1}^k\hat \phi_{l_{u_i}}-\prod_{i=1}^k\phi_{l_{u_i}}} + \sum_{\substack{v \subset u\\ v \neq \emptyset}}\sum_{\bolds{l_v}\in \disc 1{L}^{|v|}}\normemp{ \lambda_{\bolds{l_v},\bolds{l_u}}^{n} \hat \phi_{\bolds{l_v}} -\lambda_{\bolds{l_v},\bolds{l_u}} \phi_{\bolds{l_v}}}\\
&&+ \abs{C^{n}_{\bolds{l_u}}-C_{\bolds{l_u}}}.
\end{array}
\end{equation}

First, we show that
\beq\label{cvdifprod}
\left\{
\barr{c}
\alsur{\norm{\prod_{i=1}^k\hat \phi_{l_{u_i}}-\prod_{i=1}^k\phi_{l_{u_i}}}}{0},\quad \forall~l_{u_i}\in \disc{1}{L}\\
\alsur{\normemp{\prod_{i=1}^k\hat \phi_{l_{u_i}}-\prod_{i=1}^k\phi_{l_{u_i}}}}{0},\quad \forall~l_{u_i}\in \disc{1}{L}.
\earr
\right.
\eeq
Each univariate function $\hat \phi_{l_{u_i}}$ is constructed from $(\phi_{k})_{k \leq l_{u_i}}$ by the Gram-Schmidt procedure. Thus,

\beqe
\barr{lll}
\norm{\prod_{i=1}^k\hat \phi_{l_{u_i}}-\prod_{i=1}^k\phi_{l_{u_i}}} &=& \norm{\prod_{i=1}^k\left[ \frac{\phi_{l_{u_i}}-\sum_{k=0}^{l_{u_i}-1} \langle \phi_{l_{u_i}},\hat \phi_{k}\rangle_{\textrm{\tiny n}}\cdot \hat \phi_{k}} {T_n^{l_{u_i}}} \right]-\prod_{i=1}^k\phi_{l_{u_i}}}\\
&\leq & \norm{\prod_{i=1}^k\phi_{l_{u_i}} \left(\prod_{i=1}^k \frac 1 {T_n^{l_{u_i}}}-1 \right) }\\
\\
&& +\sum_{\substack{s+t=k\\ (s,t)\in \mathbb N\times \mathbb N^*\\
1\leq i_1<\cdots <i_s\leq k\\
1\leq j_1<\cdots< j_t \leq k}}\sum_{k=0}^{l_{u_{j_1}}-1} \cdots
\sum_{k=0}^{l_{u_{j_t}}-1} \norm{a_{i_1}\cdots a_{i_s}\cdot b_{j_1}\cdots b_{j_t}}
\earr
\eeqe
 where $a_{i}= \phi_{l_{u_i}}/T_n^{l_{u_i}}$, and $b_{j}=\langle \phi_{l_{u_j}}^j,\hat \phi_{k} \rangle_{\textrm{\tiny n}}\cdot \hat \phi_{k}/T_n^{l_{u_j}}$. As already proved, for all $i \in \disc 1 p$, $l_{u_i}\in \disc 1 {L}$, 
 $$
 \alsur{T_n^{l_{u_i}}} 1,~ \alsur{T_n^{l_{u_j}}} 1.
 $$
  Also, we previously showed that 
  \beqe
\alsur{  \norm{ \psemp{\phi_{l_{u_j}}}{\hat \phi_{k}} \hat \phi_{k}^j}} 0, \quad\alsur{  \normemp{ \psemp{\phi_{l_{u_j}}}{\hat \phi_{k}} \hat \phi_{k}}} 0,~~ \forall~j, ~\forall~l_{u_j}.
  \eeqe
  
Thus, we conclude that (\ref{cvdifprod}) is satisfied.\\

Secondly, as $\alsur{\normemp{\prod_{i=1}^k\hat \phi_{l_{u_i}}-\prod_{i=1}^k\phi_{l_{u_i}}}}{0}$, Assertion (\ref{item2ii}) of Lemma \ref{res1} can be applied. Assertion (\ref{itemiii}) claims that $A^{u,n}$ tends to the theoretical matrix $A_\phi^u$.
 Also, by (\ref{item2ii}) of Lemma \ref{res1}, $\alsur{D^{\bolds{l_u},n}}{D^{\bolds{l_u}}}$. Hence, $\alsur{\Lambda^{u,n}}{\Lambda^u}$. We also deduce that $\alsur{C^{n}_{\bolds{l_u}}}{C_{\bolds{l_u}}}$.\\
   
  Consequently, by induction, we deduce that every piece of the right-hand side of (\ref{systdegue}) (respectively (\ref{systdeguebis})) tends to $0$, so is $\norm{\hat \phi_{\bolds{l_u}}-\phi_{\bolds{l_u}}}$ (resp. $\normemp{\hat \phi_{\bolds{l_u}}-\phi_{\bolds{l_u}}}$).  Hence, $(\mathcal H_k)$ is satisfied.

\end{itemize}

\bigskip

As a conclusion,  $\alsur{\norm{\hat \phi_{\bolds{l_u}}-\phi_{\bolds{l_u}}}}{0}$, $\forall~\bolds{l_u}\in \disc{1}{L}^{|u|}$, $\forall~u \in S$. Hence, we deduce that 

$$
\alsur{\norm{\bar\eta_u^R-\eta_u^R}}{0},\quad \forall~u \in S,
$$ 
and
\[
\norm{\bar\eta^R-\eta^R} \leq \sum_{u \in S} \norm{\bar\eta_u^R-\eta_u^R} \Longrightarrow \alsur{\norm{\bar\eta^R-\eta^R}}{0}.
\]

\end{proof}


\begin{lemma}\label{lemcv2}
Recall that $\hat{\boldsymbol\beta}=\argmin_{\bolds\beta\in \Theta}\normemp{\mathbb Y-\mathbb X_{\hat \phi}\bolds\beta}^2$.
If (H.1)--(H.3) hold, then
\begin{equation}
\alsur{\eucl{\hat{\boldsymbol\beta}-\boldsymbol\beta_0}}{0}
\end{equation}
Moreover, 
\begin{equation}\label{lastone}
\alsur{\norm{\hat \eta^R-\bar\eta^R}} 0.
\end{equation}

\end{lemma}


\begin{proof}[{Proof of Lemma \ref{lemcv2}}]
\mbox{}

First, we remind the true regression model,
 \begin{equation}\label{model}
Y=\eta^R(\mathbf X)+\varepsilon, \quad \textrm{where}~~ \eta^R(\mathbf X)=\sum_{u\in S}  \sum_{\bolds{l_u}\in \disc{1}{L}^{|u|}}  \bolds\beta^{u,0}_{l_u} \phi_{\bolds{l_u}}(\mathbf{X_u}),
\end{equation}
with~$\mathbb E(\varepsilon)=0$, $\mathbb{E}(\varepsilon^2)=\sigma_*^2$, ~$\mathbb E(\varepsilon\cdot\phi_{l_u}(\mathbf{X_u}))=0$, $\forall~\bolds{l_u}\in \disc{1}{L}^{|u|}$, $\forall~u \in S$, and $\bolds\beta_0=(\bolds\beta^{u,0}_{\bolds{l_u}})_{\bolds{l_u},u\in S}$ the true parameter. Let 
\begin{equation}\label{mintilde}
\tilde {\boldsymbol\beta}\in \argmin_{\bolds\beta\in \Theta} \normemp{\mathbb Y-\mathbb X_\phi\bolds\beta}^2.
\end{equation}
Due to Lemma \ref{noexist}, $({}^tX_\phi\mathbb X_\phi)^{-1}$ is well defined. Thus,
\[
\tilde{\boldsymbol\beta}-{\boldsymbol\beta_0}=\left(\frac{{}^t \mathbb X_{\phi} \mathbb{X}_{\phi}}{n}\right)^{-1}\cdot \frac{{}^t \mathbb X_{\phi}\cdot \varepsilon}{n}.
\]
By the law of large numbers, $\alsur{(\frac{{}^t \mathbb X_{\phi}\cdot \varepsilon}{n})_u}{\mathbb E(\varepsilon\cdot\phi_{\bolds{l_u}}(\mathbf{X_u}))=0}$, $\forall~u\in S$. 
Moreover, $\alsur{\frac{{}^t \mathbb X_{\phi} \mathbb{X}_{\phi}}{n}}{A_\phi}$, where $A_\phi$ is defined in the new settings. Thus, $\alsur{\eucl{\tilde{\boldsymbol\beta}-{\boldsymbol\beta_0}}}{0}$.\\

Under (H.2)--(H.3), we have, by Proof of Lemma \ref{lemcv1}, that 

 \beq\label{cvfcthogs}
 \alsur{\norm{\hat \phi_{\bolds{l_u}}-\phi_{\bolds{l_u}}}}{0}.
 \eeq
 We are going to use (\ref{cvfcthogs}) to show that $\alsur{\eucl{\hat{\bolds\beta}-\tilde{\bolds\beta}}}{0}$. \\ 
 
As $\hat{\boldsymbol\beta}$ is the solution of the ordinary least-squares problem, we get $\hat{\boldsymbol\beta}=({}^t \mathbb X_{\hat \phi} \mathbb{X}_{\hat \phi})^{-1} {}^t \mathbb X_{\hat \phi} \mathbb{Y}$ because, as seen later $\alsur{{}^t \mathbb X_{\hat \phi} \mathbb{X}_{\hat \phi}}{A_\phi}$, that is invertible. \\
We define the usual matrix norm $\normmat{\cdot}$ as $\normmat{A}:=\sup_{\eucl{x}=1} \eucl{Ax}$, where $\eucl{\cdot}$ is the Euclidean norm. The Frobenius matrix norm is defined as $\normmatf{A}:=\sqrt{\mbox{Trace} (A{}^tA)}$, and $\normmat{A} \leq \normmatf{A}$~\cite{golub}. We use this inequality to get
$$
\begin{array}{lll}
\eucl{\hat{\boldsymbol\beta}-\tilde{\boldsymbol\beta}}&=&\eucl{(({}^t \mathbb X_{\hat \phi} \mathbb{X}_{\hat \phi})^{-1} {}^t \mathbb X_{\hat \phi}-({}^t \mathbb X_{\phi} \mathbb{X}_{\phi})^{-1} {}^t \mathbb X_{\phi})\mathbb{Y}}\\
&\leq & \normmat{({}^t \mathbb X_{\hat \phi} \mathbb{X}_{\hat \phi})^{-1} {}^t \mathbb X_{\hat \phi}-({}^t \mathbb X_{\phi} \mathbb{X}_{\phi})^{-1} {}^t \mathbb X_{\phi}}\cdot \eucl{\mathbb Y}\\
&= &  \sqrt{n}\normmatf{({}^t \mathbb X_{\hat \phi} \mathbb{X}_{\hat \phi})^{-1} {}^t \mathbb X_{\hat \phi}-({}^t \mathbb X_{\phi} \mathbb{X}_{\phi})^{-1} {}^t \mathbb X_{\phi}}\cdot \eucl{\frac{\mathbb Y}{\sqrt{n}}}.
\end{array}
$$

First, by (\ref{model}),
$$
\begin{array}{c}
\eucl{\frac{\mathbb Y}{\sqrt{n}}} = \eucl{\frac{\mathbb X_{\phi}{\boldsymbol\beta_0}+\varepsilon}{\sqrt{n}}} \leq  \eucl{\frac{\mathbb X_{\phi}{\boldsymbol\beta_0}}{\sqrt{n}}}+\eucl{\frac{\varepsilon}{\sqrt{n}}}\\
\end{array}
$$
and $ \eucl{\frac{\mathbb X_{\phi}{\boldsymbol\beta_0}}{\sqrt{n}}}=\alsur{\sqrt{\frac{1}{n}\sum_{s=1}^n\left[\sum_{u \in S}\sum_{\bolds{l_u}} \beta_{\bolds{l_u}}^{u,0} \phi_{\bolds{l_u}}(\mathbf{x_u}^s)\right]^2 }} { ({}^t \bolds\beta_0 A_\phi \bolds\beta_0)^{1/2}}$. Also, $\alsur{\eucl{\frac{\varepsilon}{\sqrt{n}}}}{\sqrt{\mathbb E(\varepsilon ^2)}}= \sigma_*$. Hence, $\eucl{\frac{\mathbb Y}{\sqrt{n}}}\leq ({}^t \bolds\beta_0 A_\phi \bolds\beta_0)^{1/2}+\sigma_*< \infty$.\\

Now, let us consider $ \sqrt{n}\normmatf{({}^t \mathbb X_{\hat \phi} \mathbb{X}_{\hat \phi})^{-1} {}^t \mathbb X_{\hat \phi}-({}^t \mathbb X_{\phi} \mathbb{X}_{\phi})^{-1} {}^t \mathbb X_{\phi}}$. After computation, we get
$$
\begin{array}{lll}
n\normmatf{({}^t \mathbb X_{\hat \phi} \mathbb{X}_{\hat \phi})^{-1} {}^t \mathbb X_{\hat \phi}-({}^t \mathbb X_{\phi} \mathbb{X}_{\phi})^{-1} {}^t \mathbb X_{\phi}}^2 &=& \mbox{Trace}[(\frac{{}^t \mathbb X_{\phi}\mathbb X_{\phi}}{n})^{-1}]+\mbox{Trace}[(\frac{{}^t \mathbb X_{\hat \phi}\mathbb X_{\hat \phi}}{n})^{-1}] \\
&&-2~ \mbox{Trace}[(\frac{{}^t \mathbb X_{\phi}\mathbb X_{\phi}}{n})^{-1}\cdot \frac{{}^t \mathbb X_{\phi}\mathbb X_{\hat \phi}}{n}\cdot (\frac{{}^t \mathbb X_{\hat \phi}\mathbb X_{\hat \phi}}{n})^{-1}].
\end{array}
$$
\begin{itemize}
\item  $\alsur{\frac{{}^t \mathbb X_{\phi}\mathbb X_{\phi}}{n}}{A_\phi}$ 
\item $\alsur{\frac{{}^t \mathbb X_{\hat \phi}\mathbb X_{\hat \phi} } {n} } A_\phi$, by (\ref{itemiii}) of Lemma \ref{res1}


\item $\alsur{\frac{{}^t \mathbb X_{\phi}\mathbb X_{\hat \phi}}{n}}{A_{\phi}}$  by (\ref{itemii}) of Lemma \ref{res1}.  
\end{itemize}
Under (H.1), using the result of Lemma \ref{noexist}, $A_\phi$ is invertible. Then,
\[
\alsur{n\normmatf{({}^t \mathbb X_{\phi} \mathbb{X}_{\phi})^{-1} {}^t \mathbb X_{\phi}-({}^t \mathbb X_{\hat \phi} \mathbb{X}_{\hat \phi})^{-1} {}^t \mathbb X_{\hat \phi}}^2}{\mbox{Trace}(A_\phi^{-1})+\mbox{Trace}(A_\phi^{-1})-2\mbox{Trace}(A_\phi^{-1})}=0.
\]
Thus, $\alsur{\eucl{\hat{\boldsymbol\beta}-\tilde {\boldsymbol\beta}}}{0}$. We conclude that
\[
\eucl{\hat{\boldsymbol\beta}-{\boldsymbol\beta_0}} \leq \alsur{\eucl{\hat{\boldsymbol\beta}-\tilde{\boldsymbol\beta}} + \eucl{\tilde{\boldsymbol\beta}-{\boldsymbol\beta_0}}}{0}.
\]
 At last, 
 $
 \norm{\hat\eta^R-\bar\eta^R} \leq \normmat{\mathbb X_{\hat \phi}} \alsur{\eucl{\hat{\boldsymbol\beta}-{\boldsymbol\beta_0}}} 0. 
 $
\end{proof}
 Finally, as a consequence of Lemmas \ref{lemcv1}-\ref{lemcv2}, 
 \[  
 \norm{\hat\eta^R-\eta^R} \leq\alsur{ \norm{\hat\eta^R-\bar\eta^R} + \norm{\bar\eta^R-\eta^R} } 0.
 \]

\end{document}